\documentclass{pasa}%
\pdfoutput=1
\usepackage{graphicx}

\usepackage{longtable}
\usepackage{ltablex}
\usepackage{comment}
\usepackage{threeparttable}
\usepackage{multirow}
\usepackage{rotating}
\usepackage[export]{adjustbox}
\usepackage{epstopdf}

\newcommand{\arcsec}{^{\prime \prime}}
\newcommand{\arcmin}{^{\prime}}
\def\@mcsc{ptmrc}

\title[SkyMapper change index for star formation]{SkyMapper SEDs of nearby galaxies: quenching and bursting probed by a change index for star formation}

\author[Wolf et al.]{Christian Wolf$^{1,2,3}$, Jacob Golding$^{1,2}$, Christopher A. Onken$^{1,2}$, and Li Shao$^{1,4}$
\affil{$^1$Research School of Astronomy and Astrophysics, Australian National University, Canberra, ACT 2611, Australia}
\affil{$^2$ARC Centre of Excellence for All-sky Astrophysics (CAASTRO)}
\affil{$^3$Email: christian.wolf@anu.edu.au}
\affil{$^4$Kavli Institute for Astronomy and Astrophysics, Peking University, 5 Yiheyuan Road, Haidian District, Beijing 100871, P.~R.~China}
}

\jid{PASA}
\doi{10.1017/pas.\the\year.xxx}
\jyear{\the\year}

\usepackage{aas_macros}
\hypersetup{colorlinks,citecolor=blue,linkcolor=blue,urlcolor=blue}

\begin{document}

\newcommand{\refbf}{} 

\begin{frontmatter}
\maketitle

\begin{abstract}
The wish list of astronomers includes a tool that reveals quenching of star formation in galaxies directly as it proceeds. Here, we present a proof-of-concept for a new quenching-and-bursting diagnostic, a ``change index'' for star formation, that requires only photometric data, provided they include filters such as the violet $uv$ bands used by SkyMapper. The index responds mostly to changes in star-formation rate on a timescale of 20 to 500~Myr and is nearly insensitive to dust extinction. It works effectively to distances of 100 to 150~Mpc. We explore its application to eight example galaxies in SkyMapper DR2, including known E+A and Seyfert-1 galaxies. Owing to the degeneracies inherent in broad-band photometry, the change index can only be a qualitative indicator of changes in star-formation rate. But once the SkyMapper Southern Survey is complete, the change index will be available for every spatial resolution element of every galaxy in the Southern sky within its working distance range.
\end{abstract}

\end{frontmatter}

\section{INTRODUCTION}\label{sec:intro}

Galaxies have changed throughout the history of our Universe, a process that involves many mechanisms. Arguably the biggest transformational experience in their lives is ``quenching'', i.e. a strong suppression of their star formation \citep[e.g.][]{DG83,Peng10,Scha14}. Images of the sky show galaxies with a wide range of appearances, but in the quest for understanding their transformation one aspiration has always loomed large: to catch them in the act of changing. While astronomers cannot watch galaxies long enough to see changes directly, they work like archaeologists instead and compare specimens from different epochs to observe differences in populations and infer change events from that.

Mapping the birth rate of new stars in galaxies, a.k.a. their star-formation rate (SFR), is an established tool in astronomy \citep{Kennicutt98,Calzetti13}. So far, change in galaxies was inferred indirectly from the fact that galaxy A looks different from galaxy B; e.g. the outer parts of some spiral galaxies in the Virgo Galaxy Cluster were shown to lack young stars compared to spirals in the open galaxy field \citep{KK04a,Cortese12}. The inference is that Virgo spirals had star formation in their outer parts previously and this has stopped owing to the cluster's influence. However, it would be more desirable to have a tool that directly shows temporal changes in the birth rate of stars as they occur, as this would enable to see more immediate correlations between a galaxy's physical conditions and the occurring change.

The ideal goal is to reconstruct the star-formation history from the age mix of the present stellar population. With a purely photometric approach this requires many bandpasses from the ultraviolet and optical regime, as well as far-infrared data to break degeneracies between the ages of stars and the strength of dust extinction \citep[e.g.][]{Boselli16}. 

An alternative approach is to constrain star-formation history by combining photometry and spectroscopy \citep[e.g.][]{Scha07b,Crowl08}. Using this approach on Virgo galaxies, \citet{Crowl08} found that the outer disks of all galaxies with truncated star-forming disks appear to have been quenched less than 500~Myr ago. \citet{Fossati18} study the specific example of NGC 4330, and find a slow outside-in progression of a quenching front in star formation.

A further method involves decomposing optical spectra of galaxies into age components \citep[a.k.a. the fossil record,][]{Heavens04, GarciaBenito17}, but this approach reveals only the mean behaviour of large samples and does not work for individual galaxies.

Currently, the main diagnostic for recently quenched star formation is spectroscopic observations of a strong H$\delta$ absorption line together with an absent [O{\sc ii}] emission line \citep{DG83, CS87}. This signature is created by the absence of stars younger than 10~Myr, plus a clear presence of A-type stars with a mean lifetime of 300~Myr. 

Early work on searching for quenching signals was mostly based on spectra from the first stage of the Sloan Digital Sky Survey \citep[SDSS,][]{York00}, which cover both lines together only in galaxies at over 150~Mpc distance, and whose single fibres probe only galaxy cores and lack spatial resolution; hence, all quenching found with SDSS has been in galaxies with quenched cores at over 150~Mpc distance \citep{Goto03}. However, the star-formation properties measured in the inner parts of a galaxy can differ systematically from their outer parts \citep{Richards16}.

An ideal diagnostic, of course, maps galaxies with spatial resolution to enhance sensitivity to localised changes. Among galaxies in the Virgo Cluster, e.g., there are many cases where star formation proceeds undisturbed in the inner parts, while their outer disks are devoid of gas, dust and star formation \citep{KK04a, Cortese12,Fossati18}. The time scales on which star formation in these galaxies is quenched have been estimated to range from 100 to 500~Myr \citep{Boselli16}. \citet{Fossati18} argue that local quenching happens usually on shorter timescales while global quenching of a whole galaxy may take much longer. If observed with a single fibre, all those local signals will be blended with the rest of the galaxy, lowering contrast and inhibiting their detection. Spatial resolution has been key to discern the otherwise diluted signals of transformation.

Modern spectroscopic integral-field units (IFUs) can map spectra of galaxies across several resolution elements, allowing a more holistic study of galaxies. These include e.g. the Potsdam Multi Aperture Spectrophotometer instrument \citep[PMAS/PPak,][]{Roth05,Kelz06} at the Calar Alto 3.5m-telescope, the Sydney-AAO Multi-object Integral-field spectrograph \citep[SAMI,][]{Croom12} at the 3.9m Anglo-Australian Telescope (AAT) and the Mapping Nearby Galaxies at APO \citep[MaNGA,][]{Drory15} instrument. Indeed, the CALIFA Survey at Calar Alto \citep{Sanchez12} and the SAMI Survey \citep{Bryant15} have observed about 600 and 3 000 galaxies, respectively, and MaNGA \citep{Bundy15} will observe the internal structure in about 10 000 galaxies in unprecedented detail. However, the field-of view of IFUs is arguably limited and either requires more distant galaxy targets or restricts the data to central regions in a galaxy.

In this paper, we present a proof of concept for a new diagnostic tool that can map temporal changes in the birth rate of stars. It targets a similar physical phenomenon as H$\delta$-line spectroscopy given that it measures the relative weight of A-type stars in the population. Its main advantage is that it is available as full-format imaging data. It uses optical spectral passbands within the sensitivity range of common CCDs, and is in principal available at every telescope with imaging cameras as long as suitable filters are provided. By relying on filters the principal limitation is that the method works only in a narrow redshift interval, where the crucial spectral features of a stellar population are aligned with the filter transmission curves.

We appreciate that our new method uses far less information than spectroscopy provides and in this first step also ignores the wide range of multi-wavelength data available in modern astronomy. The reader may thus suspect that there will be great degeneracy within the high-dimensional space of possible star-formation histories and galaxy properties. We deliberately avoid large complexity in both data and analysis and intend to clarify in this paper what optical filters alone can tell us. In this spirit, we also do not employ a black box-style SED fitting analysis, but try instead to extract which effects are expected to stand out from the multitude of degeneracies. 

As part of this effort, we have devised a particular diagnostic that minimises sensitivity to dust extinction and general ageing of stars. Since the amount of dust varies greatly within and among star-forming galaxies, as does the age distribution of stars within a galaxy, any diagnostic of short-term change must be insensitive to dust and secular ageing trends to be truly useful. We construct an SFR change index from the passbands of the SkyMapper Southern Survey \citep{Wolf18} and show that SkyMapper filters have a unique advantage over SDSS filters, whose broad-band spectral energy distributions (SEDs) do not provide such a diagnostic. Essentially, this new index probes the age mixture among young stars in populations seen at redshift $z<0.03$ or distances up to $\sim 130$~Mpc.

The SkyMapper Data Release 2 currently provides the largest data set in these filters \citep{Onken19}. When the Southern Survey is complete, it will provide data for over 10 000 large and well-resolved galaxies within 100~Mpc distance, covering the entire Southern Hemisphere. Currently, we know less about many of the galaxies within 100 Mpc distance than about some of those at larger distances. The SAMI Survey, e.g., observes 3 000 galaxies all with distances beyond 100 Mpc because of such field-of-view limits. Hence, SAMI and the potential SkyMapper sample are targeting complementary volumes with little overlap. At the nearby end of the distance scale there are also efforts to reconstruct the star-formation history from observations of resolved stellar populations such as the ACS Nearby Galaxy Survey Treasury \cite[ANGST,][]{Dalcanton09}, which reaches out to 4~Mpc distance. While these observations are not deep enough to reach stars of a few hundred Myr age on the main sequence, they reveal sufficiently luminous Helium core-burning stars of that age; these hold clues on their galaxies' recent star-formation history although mismatches with models currently limit our ability to interpret them \citep{McQuinn11}. 

This paper presents our new change index based on SkyMapper filters. In Sect.~2 we devise the method using stellar population synthesis, and discuss its dependence on metallicity, redshift and observing conditions. We also investigate the role played by AGN and foreground stars in limiting the interpretation of changes in star formation. In Sect.~3 we show first examples of observed signatures in the SkyMapper Southern Survey, including AGN and foreground stars. We discuss current limitations of the method in Sect.~4 and close with an outlook in Sect.~5. Throughout the paper, magnitudes from 2MASS are in the Vega system, while $uvgriz$ magnitudes are in the AB system \citep{Oke83}. We adopt a Hubble-Lema\^itre constant of $H_0 = 70$~km~s$^{-1}$~Mpc$^{-1}$.

\section{THE CHANGE INDEX IN BIRTH RATE}\label{method}

\subsection{From stars to galaxies}

Our new diagnostic for changes in the birth rate of stars is a combined bursting-and-quenching diagnostic. It is currently available at the SkyMapper Telescope, where it is a pleasant by-product of a feature designed for Galactic archaeology studies. When SkyMapper was planned, the Sloan Digital Sky Survey had already revolutionized galaxy surveys by imaging much of the Northern sky. When SkyMapper filters were designed, their scientific niche was understanding the stars of the Milky Way. 

The biggest difference between the SkyMapper and SDSS filters is on the blue side of 400~nm: in place of the SDSS $u$ band with $(\lambda_{\rm cen}/{\rm FWHM})=(358{\rm nm}/55{\rm nm})$ SkyMapper has a violet $v$ band (384/28) and a more ultraviolet $u$ band (349/42). This choice of filters adds direct photometric sensitivity to the surface gravity of stars via the strength of the Hydrogen Balmer break driving changes in the $u-v$ colour, as well as to their metallicity via metal lines affecting the $v-g$ colour. By exploiting this information, SkyMapper has led to the identification of a large sample of extremely metal-poor stars (Da Costa et al., in preparation) including the most chemically pristine star currently known \citep{Keller14}. In this paper, however, we explain the use of this filter pair as a bursting-and-quenching diagnostic.

Our primary idea was that filters that are useful for characterising stars are also useful for characterising stellar populations in galaxies at redshifts close to zero. In more distant galaxies the spectral energy distributions (SEDs) of stellar populations are redshifted out of sync with the passbands. We first note that the SkyMapper colour index $u-v$ takes on an extremely red value for main-sequence A-type stars. Assuming a fixed continuum colour redwards of 400~nm, the $u-v$ of a stellar population then depends on the fraction of A-type stars in the mix. After a quenching event, the young O- and B-type stars die away and leave the blue side of the galaxy SED dominated by the bright $(u-v)$-red A-stars. Conversely, in the event of a starburst a sudden increase in the fraction of O- and B-stars above typical levels makes the $u-v$ index bluer than usual. A continuously star-forming and steadily evolving population, however, has a constant O-to-A star ratio and will thus show a constant $u-v$ index, irrespective of the detailed history of an older underlying population. 

In the following, we explore how the $u-v$ colour index reflects changes in the star-formation rate and what other factors it depends on. We start with trying to understand the primary signal as it originates from the stellar population, and only step by step consider factors that alter a galaxy's appearance, such as dust reddening, nebular emission lines, the specific effects of Active Galactic Nuclei (AGN), and finally the effect of Galactic foreground stars.

\begin{table}
\caption{SkyMapper filters and DR2 imaging data. FWHM is for the median PSF. $\Sigma$ limit is the surface-brightness sensitivity of a single Main Survey exposure at 3~sigma. }
\label{filtab}      
\centering          
\begin{tabular}{llcccc}
\hline\hline       
Filter & $\lambda_{\rm cen}$/$\Delta \lambda$ & FWHM & $\Sigma$ limit  \\ 
	   & (nm) & (arcsec) & (ABmag) \\
\hline
$u$ & 349/42  & $3.1$ & 22.0  \\ 
$v$ & 384/28  & $2.9$ & 22.1  \\ 
$g$ & 510/156 & $2.6$ & 23.9  \\ 
$r$ & 617/156 & $2.4$ & 23.6  \\ 
$i$ & 779/140 & $2.3$ & 22.6  \\ 
$z$ & 916/84  & $2.3$ & 21.8  \\ 
\hline                  
\end{tabular}
\end{table}

\subsection{Stellar population synthesis}

We first develop the change index using predictions from stellar population synthesis. We calculate spectral energy distributions (SEDs) in a range of passbands including the SkyMapper and SDSS filters for a grid of possible star-formation histories. Our observational targets are galaxies in the local Universe, which usually include a substantial amount of old or intermediate-age stars. We aim to detect quenching and bursting events in mature galaxies, long after most of the stars in a galaxy have formed. 

We model our stellar populations with exponential star-formation histories proceeding for 10~Gyr, before a quenching or burst event happens. Immediately following the event, the SEDs of the populations change rapidly and hence we desire great time resolution in calculating the SED evolution. Because of this, we use the code {\it Stochastically Lighting Up Galaxies} (SLUG) by \citet{SLUG} that provides freedom to set time steps for the calculation. 

A second interesting aspect of SLUG is that it follows the evolution of every single star in the population and thus incorporates the stochasticity of a finite stellar mass, in terms of sampling the initial mass function (IMF). SkyMapper data provide resolution elements smaller than 100 pc in size for the nearest galaxies. In principle, SLUG allows us then to find out how much noise stochasticity might cause in subtle aspects of the SED in the low surface-mass density outskirts of galaxies, however, we will not exploit this aspect in our first analysis.

\begin{figure}
\begin{center}
\includegraphics[angle=270,width=\columnwidth,clip=true]{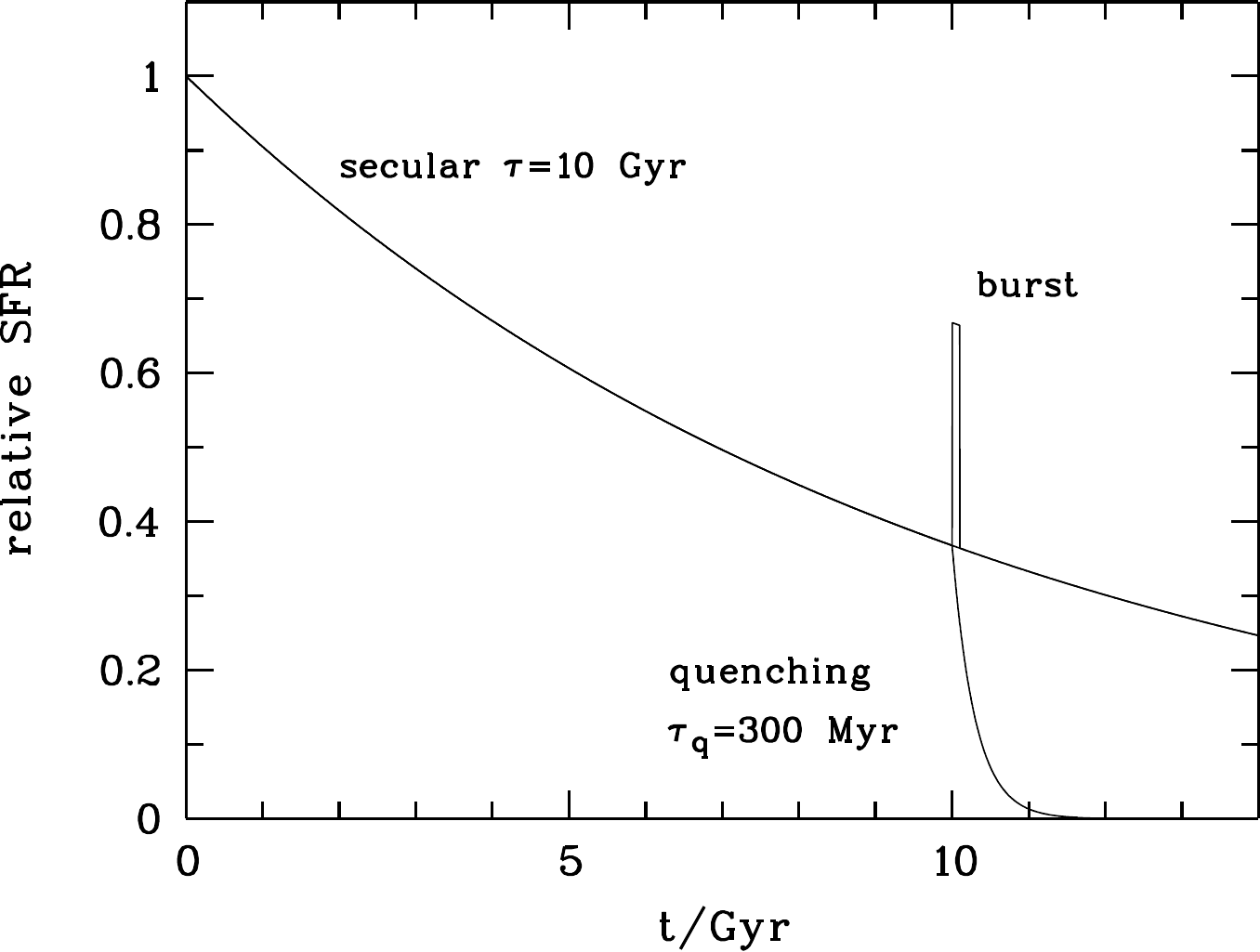}
\caption{Three classes of star formation histories (SFH) are defined in our parametrisation: (1) a secular evolution without any special event, (2) an SFH permanently quenched at some point, and (3) an SFH with a temporary burst. All cases are shown for an underlying secular SFH with a decline time scale of $\tau=10$~Gyr.}\label{SFH_example}
\end{center}
\end{figure}

\subsubsection{Basic SLUG runs}

Given that SLUG follows every single star in evolving a population, it is very compute-intensive. Also, we wish to investigate a broad range of star formation histories (SFHs) including quenching and bursting events within mature populations after a long time of secular evolution. The effects of these events on the galaxy SED need to be considered with high time resolution on the order of 10~Myr, which makes regular use of SLUG extremely expensive. Hence, we use a shortcut that we call the Impulse Response Method. In essence, we reverse-engineer existing functionality that may appear redundant, but we increases computational speed.

We first run SLUG with a 5~Myr short time bin of constant star formation rate to describe an elementary building block of a star-formation history with a total stellar mass of $10^5{\rm M}_\odot$. We sample the SED evolution of the building block at a high time resolution of 10~Myr for a duration of 14~Gyr. In a second step, we combine the building blocks of a more complex star formation history into a full SED. Using the discrete data from SLUG for the flux response to the short star burst $F_b(t)$ and a given SFH $H(t)$, the total flux $F(t)$ is
\begin{align}
F(t)=\sum_{t'=0}^tF_b(t')H(t')  ~.
\end{align}
The major downside of our method relative to the design purpose of SLUG is that we lose continuous calculations of the metallicity in the simulated environments (metallicity effects are discussed in detail in Section \ref{sec:Metallicity}).

We have chosen to consider at first only the model spectra for the stellar population and ignore the effect of emission lines, since their appearance depends with great sensitivity on the geometry of a galaxy. Their effects are considered separately in Section~\ref{Elines}. 

\subsubsection{Multi-parameter grids}

In order to investigate the space of possible SFHs, we first define a parametrisation of potential SFHs. As mentioned above we use exponential $\tau$ models to describe a normal stellar population with SFR described by 
\begin{equation}
f(t)=Ce^{\frac{t}{\tau_i}}   ~,
\end{equation}
where $f$ has units of mass over time, $t$ is a measurement of elapsed time, and $C$ is a scaling constant. We investigate the sensitivity of the change index based on varying initial values for $\tau_i$, those being $\tau_i=0.3,\ 1,\ 3$ and $10$~Gyr. Also included is a $\tau_i=\infty$ case, representing a constant SFR. 

At $t=10$~Gyr we introduce an event that is either a star burst or a quenching event. Quenching events are parametrised by a second tau model with $\tau_q=0.03,\ 0.1$ and $0.3$~Gyr. We also include a case of complete quenching, $\tau_q=0$. Burst events are represented as a 100~Myr period of constantly increased star formation. They are parametrised by a 1, 3, 10, 30, or 100\% increase of the instantaneous SFR at $t=0$~Gyr. We compare the response of the change index to the varying event parametrisations, and also include a case where the initial star formation continues unchanged after $t=10$~Gyr. A graphical interpretation of this parametrisation of SFH is shown in Figure~\ref{SFH_example}.

\begin{figure*}
\begin{center}
\includegraphics[angle=0,width=0.9\textwidth]{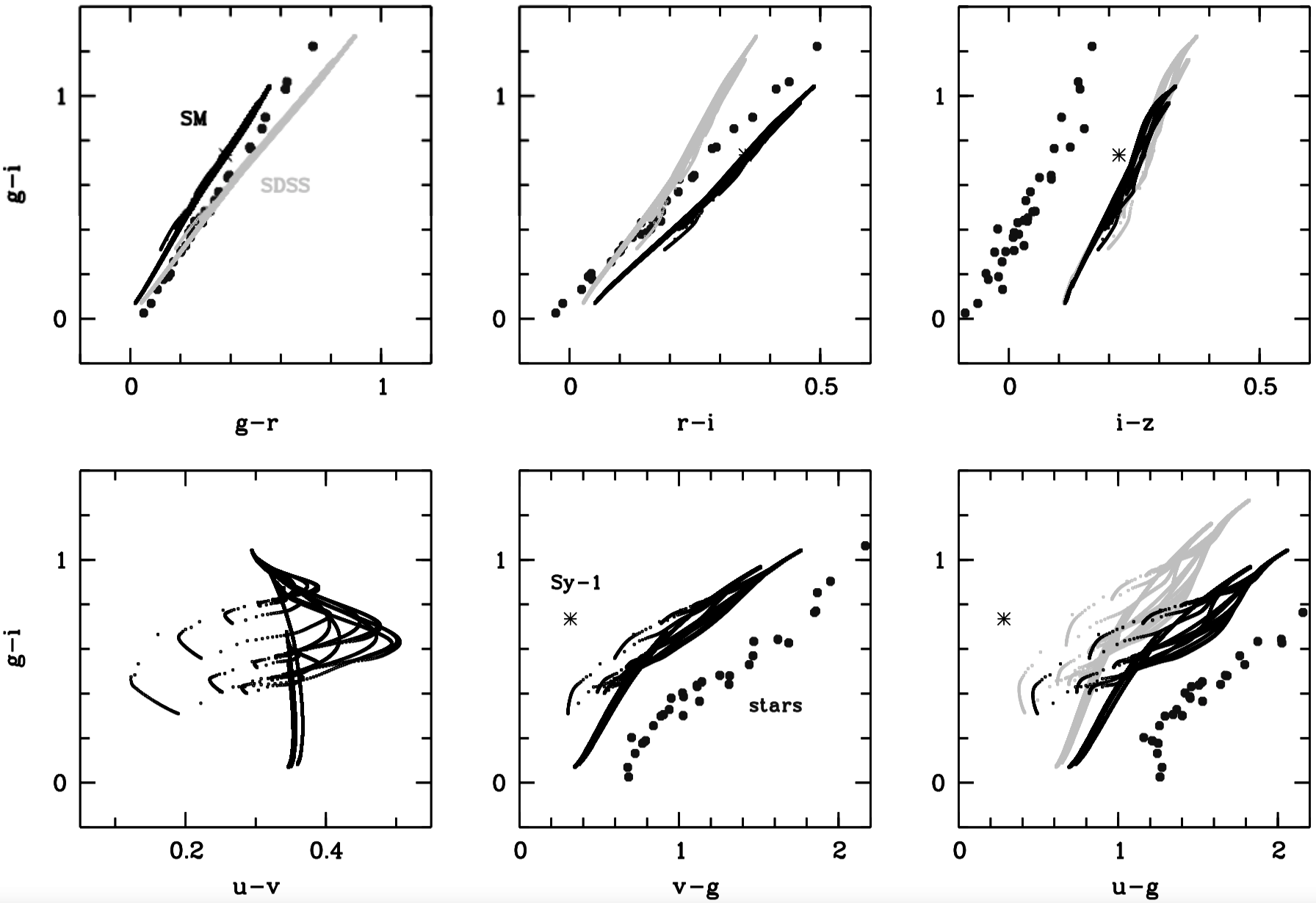}
\caption{Colour-colour tracks in SkyMapper (SM, black) and SDSS (grey) passbands for a variety of evolutionary scenarios including secular evolution for 14~Gyr, quenching and starbursting events. SkyMapper and SDSS passbands are different but have in common that stellar populations form a 1-parameter family in $griz$ (top row). In both filter sets, the $u$ band adds a second dimension (bottom right), and in SkyMapper the bespoke $v$ band adds a third one. The asterisk marks a mean Seyfert-1 SED, and dark grey symbols are stars. }\label{CC}
\end{center}
\end{figure*}

\subsubsection{Evolutionary tracks in SkyMapper passbands}

First we investigate the evolutionary tracks of our stellar populations assuming no dust reddening. 
We consider three basic scenarios: constant star formation, a sudden shutdown in star formation after 10~Gyr, and a burst after 10~Gyr, where star formation doubles for the duration of 100~Myr. Passband fluxes for the latter two respond with a symmetric deviation from the constant scenario during the first 100~Myr, while colours respond asymmetrically due to their nature as ratios of two differently evolving variables. Towards redder passbands fluxes respond more slowly; colours may get monotonically redder during quenching as in the case of $g-i$ or revert to their original value as in the case of $u-v$.

Figure~\ref{CC} shows a complete set of colour-colour diagrams for the evolution of galaxy SEDs in the filter sets of SkyMapper and SDSS. It includes a range of secular $\tau$-values from 3 to 100~Gyr as well as quenching and bursting options, over a time range from 1 to 14~Gyr after formation. For clarity, only unreddened populations are shown. Shifts between the loci of the SkyMapper (labelled SM) and SDSS colours reflect the differences among the passbands. Also, the $v$ filter exists only in the six-filter set of SkyMapper.

In the top row, we see that the passbands $griz$ span effectively only a 1-parameter family of population colours, where time since formation, the $\tau$ of secular evolution, and the strength of and phase within a quenching or bursting event are all degenerate. Even dust reddens the SEDs along the 1-parameter relation found here. We choose to represent this single parameter by $g-i$; $g-z$ has a wider separation in wavelength and thus a larger signal, but $z$ band data are generally noisier than $i-$band data. Of course, a full SED analysis may include redundant filters to reduce noise.

In the bottom row, we show the additional information provided by filters on the blue side of 400~nm. SDSS has a single $u$ band, so the single 2-D diagram of $g-i$ vs. $u-g$ (bottom right in Figure~\ref{CC}) contains all discernible information on the SFH of a galaxy. SkyMapper, in contrast, offers an additional dimension of information due to the $v$-filter. In the bottom left panel of Figure~\ref{CC} we see that SkyMapper $u-v$ has hardly any dependence on the secular $\tau$ or the overall mean age and time since formation; instead $u-v$ responds distinctly to quenching and bursting events, which makes it a candidate for an indicator of recent change in SFR. Pending an investigation of dust effects, it is conceivable that the SkyMapper filter set can disentangle the effects of mean age of the population and recent changes in the star formation history, while SDSS passbands offer too few constraints to disentangle these two parameters.

\subsection{A change index with minimal sensitivity to dust reddening and secular ageing}

Across the face of a galaxy, the amount of dust can vary greatly and introduce colour changes into the SED that do not stem from star-formation changes. Thus, a diagnostic of star formation must be insensitive to dust to be truly useful. Generally, populations of stars appear redder, when they are older or more reddened by dust. As we have no {\it a-priori} information about age or dust, we need to estimate this additional reddening from other passbands. In the following, we describe step by step how we arrived at our change index.

\subsubsection{Constructing insensitivity to dust and age}

In the absence of dust, the SkyMapper $u-v$ colour has virtually no sensitivity to time during any slowly varying star-formation history. For all ages from 1 to 14~Gyr at $\tau>5$~Gyr it has a mean AB colour of 0.35~mag and a peak-to-peak colour range of 0.013~mag. However, it responds quickly to short-term variations in star formation: assuming an underlying population that has formed stars for 10~Gyr with $\tau=10$~Gyr, a sudden doubling of star formation causes the $u-v$ colour to become bluer by 0.05~mag after 25~Myr, while a sudden shutdown of star formation makes it redder by 0.1~mag within the same time. We calibrate $u-v$ as a change index (CI) with negative values for bursts and positive values for quenching by subtracting its mean value for secular evolution and thus get 
\begin{equation}
  CI_{v1} = (u-v)-0.35   ~.
\end{equation}
Dust reddening will change all colour indices including $u-v$ as prescribed by the reddening law. As a fiducial, we adopt a \citet{F99} law with $R_V=3.1$ and make the change index insensitive to dust of this kind by estimating reddening from $g-i$, using simply the reddening law to subtract a scaled dust contribution. In our case a 1~mag reddening in $g-i$ corresponds to a 0.19~mag reddening in $u-v$; thus we get
\begin{equation}
  CI_{v2} = (u-v)-0.19(g-i)-0.35   ~.
\end{equation}
This correction ensures that all evolutionary tracks land on the same CI irrespective of $E(B-V)$. However, by including a $g-i$ term we have imported an undesired age sensitivity into the CI that we would like to remove without reintroducing dust sensitivity. 

\begin{figure*}
\begin{center}
\includegraphics[width=0.9\textwidth,clip=true]{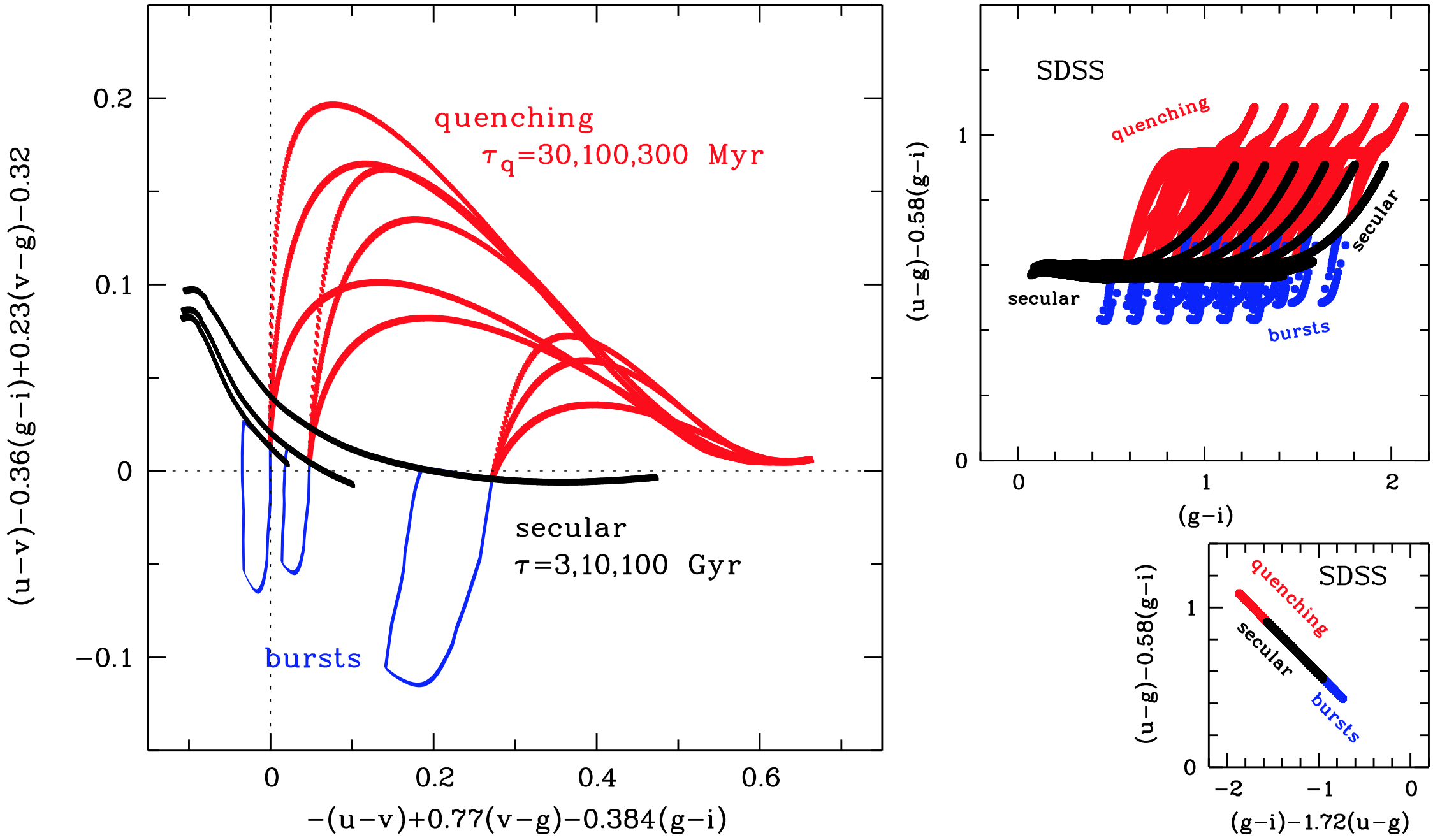}
\caption{Left: The SkyMapper change index (y-axis) confines age effects of secular evolution (black points) in a narrow interval. Quenching and bursting events (red and blue trajectories) produce positive and negative signals. Lines for different $E(B-V)$ are indistinguishable as this diagram has no reddening vector. Right: A similar index built from SDSS filters shows broad ambiguities, and dust-dependence on at least one axis. Removing the dust dependence on both axes creates a trivial 1-D plot (bottom right).} \label{CI_SMSDSS}
\end{center}
\end{figure*}

To this end, we create an age index by studying the curvature of the SED from the $vgi$ passbands, using 
\begin{equation}
  AI_{v1}=(v-g)-(g-i)=v-2g+i	~.  
\end{equation}
This age index can be made formally dust-insensitive using a correction term built from the reddening law again, resulting in
\begin{equation}
  AI_{v2}=(v-g)-0.744(g-i)	~.  
\end{equation}
Next, we scale this age index suitably and add it to the change index to minimise residual age sensitivity; we find that adding $+0.23\times AI_{v2}$ achieves this goal, giving $CI_{v3}=CI_{v2}+0.23\times AI_{v2}$. When plotting $CI_{v3}$ vs. $AI_{v2}$ we notice that quenching and bursting events seem to move the population along diagonal lines, and we skew the diagram by defining $AI=AI_{v2}-CI$, which makes bursts and quenching events appear vertical. This final diagnostic diagram for SkyMapper is thus defined by 
\begin{align}
  CI&=+(u-v)-0.360(g-i)+0.23(v-g)-0.32 & \nonumber \\
  AI&=-(u-v)-0.384(g-i)+0.77(v-g) &
\end{align}
and is shown in Figure~\ref{CI_SMSDSS}. The evolutionary tracks in the figure include three cases of secular evolution (black lines) with $\tau$ value of 3, 10 and 100~Gyr, where the first, most steeply declining one, extends across the widest range in AI changes. From the 10~Gyr point in time on each secular track extends a burst track downwards. Only one burst amplitude is shown for clarity; weaker bursts draw smaller clock-wise loops and bursts of extreme strength saturate around $CI\approx -0.3$. Upwards extend three quenching curves each, corresponding to nearly sudden ($\tau_q=30$~Myr) or slower ($\tau_q=100$ and 300~Myr) quenching. Steeper quenching makes stronger signals.

The axes in this diagram are not orthogonal, but they are intuitive to interpret; the x-axis shows effects of time and age:

\begin{enumerate}
\item Time and secular evolution move populations to the right unless the SFR increases, and they do so more quickly when $\tau$ is shorter.
\item Bursts move them to the left, and quenching events quickly move them to the right. 
\end{enumerate}

More important for us is the y-axis, which shows short-term changes in SFR:
\begin{enumerate}
\item Secular evolution with only slow changes in SFR keeps populations close to the zero level.
\item Quenching events quickly move them up, while bursts quickly move them down. 
\end{enumerate}

By design, this figure is dust-insensitive for $R_V=3.1$ and indeed shows the whole grid from $E(B-V)=0$ to 0.5. However, changes in the reddening law will cause the evolutionary tracks to move around. As an alternative, we considered the significantly different extinction law given by equations 3 and 4 in \citet{Calzetti00} with their chosen value of $R^\prime_V = 4.08$. This choice changes the reddening curve to $R_u-R_v=0.488$, $R_v-R_g=1.331$, $R_g-R_i=1.719$, and as a result adds to the change index the term $\Delta CI_{\rm Calzetti}=-0.098 (g-i)$.

We can see that the secular tracks produce some positive CI signals at young ages, where the CI can see the beginning of the history in the stellar population, and the mix of star types has not yet converged to a long-term average. These positive values do not signal quenching events, but they will also not be mistaken as such since they appear only at negative AI values. As soon as $AI>0$, any positive CI value unambiguously indicates quenching without any cross-talk from secular ageing or dust. The zeropoint of the x-axis represents a 10~Gyr old population with constant SFR to within 0.01~mag. It appears that partial quenching, leading only to small positive values, would remain undetected if it occurred in a young population to the left of $AI=0$. However, the upper one of the three secular tracks represents $\tau=3$~Gyr, and any secular curve above it would need to represent a faster SFR decline essentially representing quenching itself. Finally, negative values of CI unambiguously represent bursts at all times.

\subsubsection{Comparison with SDSS alone} 

Figure~\ref{CI_SMSDSS} shows SDSS colour indices on the right. We started with $u-g$ as a change index and $g-i$ as an age index. We then made the first index formally insensitive to dust, yielding $CI_{\rm SDSS}=(u-g)-0.58(g-i)$. The only way to make the second index dust-insensitive produces an index that is linearly dependent on $CI_{\rm SDSS}$; this produces a trivial one-dimensional plot, in which dust plays no role, but ageing, bursting and quenching are fully degenerate (see bottom right panel).

\subsubsection{Including ultra-violet passbands}

An alternative approach combines SDSS-like filters with the ultraviolet passbands from the {\it Galaxy Evolution Explorer} \citep[$GALEX$,][]{Martin05}. This works well in the absence of dust, and thus in particular in the outer parts of galaxies where the dust column density is low, although \citet{George19} apply it also to study quenching in the bar of a galaxy.

However, each 0.1~mag step in the $E(B-V)$ level of dust extinction will on average halve the light output in the $GALEX$ FUV and NUV bands, which makes it difficult to estimate the fraction of young stars in the presence of dust. Indeed, \citet{Boselli16} were able to use $GALEX$ and optical data for studying quenching in Virgo galaxies after combining it with far-infrared data that estimate the dust effects from its warm emission.

\citet{Scha14} have also used a combination of $GALEX$ and SDSS photometry of galaxies, albeit without far-infrared data, to identify two separate quenching scenarios that correlate with galaxy morphology and thus presumably the physical quenching mechanism. However, they used integral photometry instead of spatially resolved galaxy maps, and as long as there are some UV-emitting stellar populations outside the volume screened by dust, differences will be observable. In our case, we are looking at maps of galaxies and the known strong variations of dust extinction across the face of a galaxy will introduce strong spatial variations into the UV-optical colour maps.

Also, when populations with different levels of dust extinction are mixed within a resolution element, the more extinguished ones will contribute little to the UV light; they will then fail to contribute to the average flux in $GALEX$ bands and thus bias the UV-optical colours. In contrast, the $u$/$v$ filter pair in SkyMapper is so close in wavelength space that the dust term between them is small. Hence, differently extinguished parts of the population will contribute to both filters with nearly equal weight and not bias the change index much.

\begin{figure}
\begin{center}
\includegraphics[angle=270,width=\columnwidth]{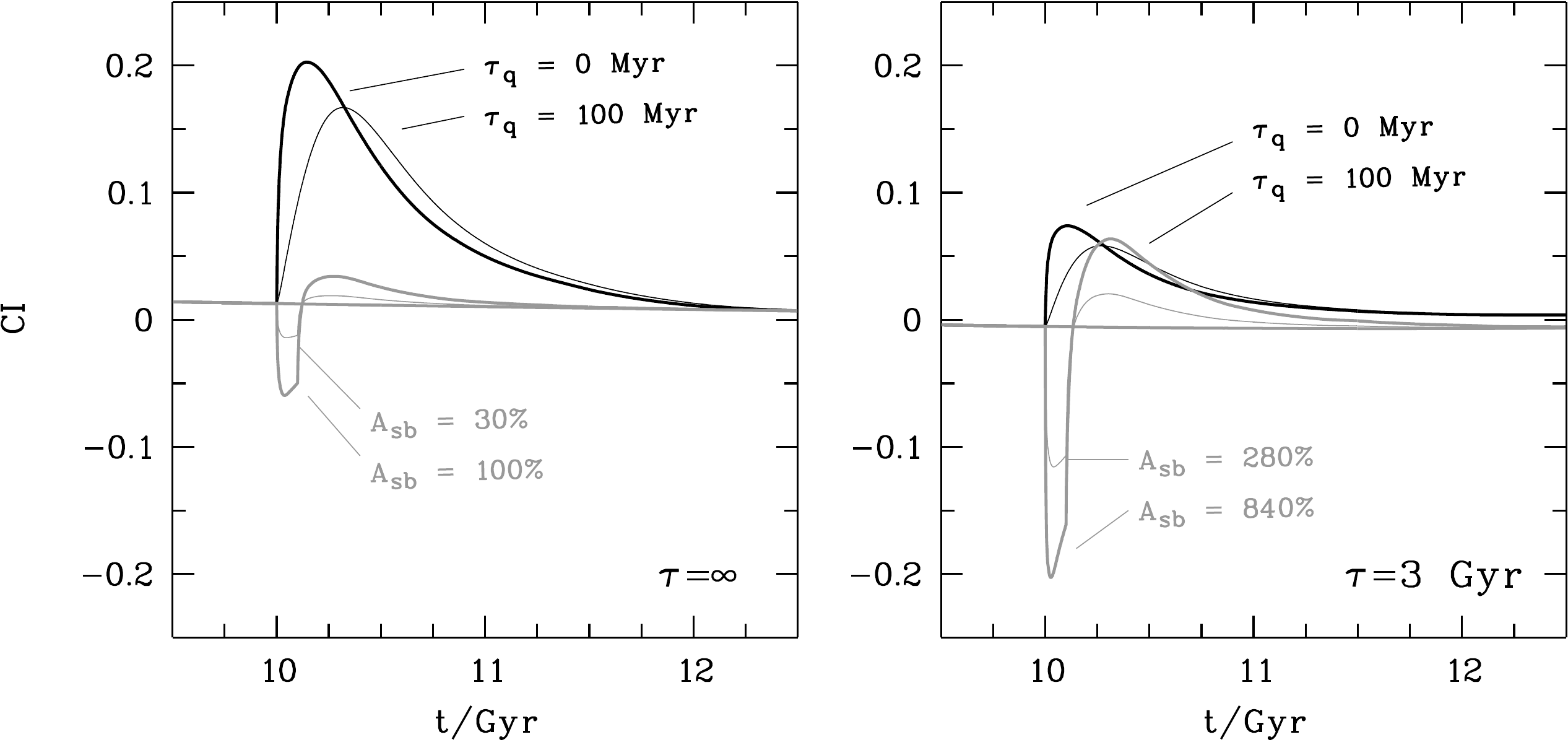}
\caption{Time evolution of the change index: an event happens 10~Gyr after a galaxy forms, following either constant star formation (left panel) or an exponential decline with $\tau=3$~Gyr (right panel). Burst events lasting for 100~Myr are described by an increase $A_{\rm sb}$ in SFR above the current rate and lead to negative index values. Quenching events produce positive signals and are described by a quenching timescale $\tau_{\rm q}$.
}\label{CI_timeevo}
\end{center}
\end{figure}

\begin{figure*}
\begin{center}
\includegraphics[width=\textwidth,clip=true]{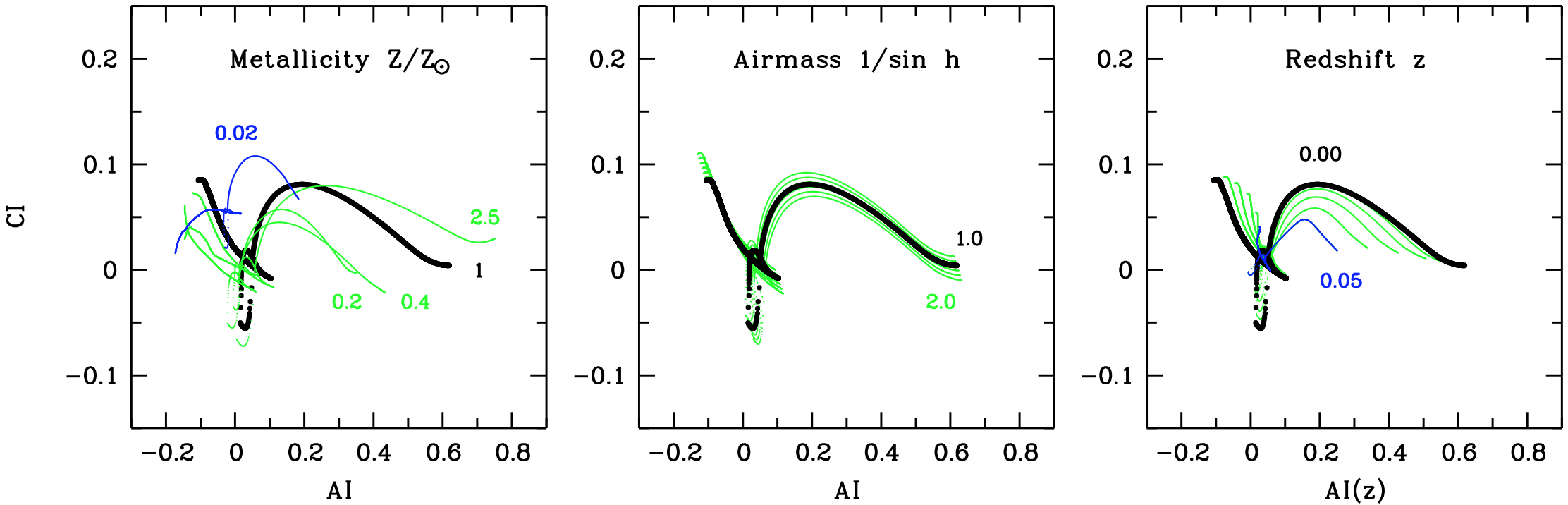}
\caption{Change index dependence on metallicity of the stellar population (left), airmass of observation (centre) and redshift (right). We show tracks for $\tau=10$~Gyr, and only one quenching and one bursting event. Black tracks denote solar metallicity, redshift zero and airmass 1. Redshift and airmass of observation are generally known and can be corrected, but metallicity is {\it a-priori} unknown and tends to change across a galaxy. However, as long as metallicity is not as low as 1/50th solar, the tracks are qualitatively similar, and quenching, bursting and secular evolution can still be distinguished, although the signal amplitudes vary. } \label{CI_3EFF}
\end{center}
\end{figure*}

\subsubsection{Dependence on time}

Figure~\ref{CI_timeevo} shows the time evolution of the change index after a bursting or quenching event that follows a 10~Gyr long evolution of the galaxy. The left panel shows the case where the galaxy had a constant SFR before the event. In this situation, quenching events produce much stronger signals than bursting events. In the right panel most of the stars in the galaxy formed a long time ago, following an exponential decline with $\tau=3$~Gyr. In this case, the burst events can leave stronger signals as they are superimposed on an old population. Just after the end of the burst, the change index does not revert simply back to zero, but overshoots into positive territory indicating that the end of the burst appears like a weak quenching event. 

One of the plotted quenching scenarios involves a sudden full truncation in star formation. This can be thought of as a point response to a star-formation change event. The colour signal in our change index peaks 50 to 300 million years after a change; more complex scenarios are measured as the convolution of the evolving birth rate with this response profile. This time filtering is also the reason why the end of a burst that lasts for only 100~Myr will not appear as a strong quenching event. 

The filtering time scale corresponds roughly to the internal dynamical time scale within large galaxies. Changes in star formation that are local to one part of a galaxy might thus have spread out over larger parts of the galaxy. This will limit the contrast at which a quenching and bursting signals will persist. The steep rise of the change index signal at the onset of an event will make localised events appear with full contrast and spatial resolution, but later migration and diffusion will make the signal disappear more quickly than if the affected population remained spatially contained.

\subsubsection{Sensitivity to metallicity}
\label{sec:Metallicity}

Figure~\ref{CI_3EFF} shows how the main tracks of our change index for metallicities from 0.02 to 2.5 times solar. More metal-rich populations produce generally redder indices, but we focus on a dependence of CI on metallicity, not those of AI, as the latter are degenerate with secular effects anyway. The main uncertainty is that the location of the secular track and the amplitude of the change index caused by quenching depend on metallicity for subsolar populations. There is a factor 2 difference in the signal amplitude between solar metallicity ${\rm Z}_\odot$ and $0.2{\rm Z}_\odot$. While there is a clear degeneracy between metallicity and the strength of the quenching event, there will be no ambiguity about the direction of the effect as metallicity has little effect on the secular track\footnote{The extremely metal-poor population with 1/50th solar metallicity behaves very differently, but we don't expect to observe them in low-redshift galaxies.}.

\subsubsection{Constructing insensitivity to observing airmass}

The effective transmission curve of the SkyMapper $u$ band depends on the airmass of observation, due to two factors: 
\begin{enumerate}
\item below 350~nm atmospheric extinction increases rapidly towards shorter wavelengths such that higher-airmass observations are made through a relatively redder passband
\item the SkyMapper $u$ band has a red leak, which becomes more dominant towards higher airmass as it is less affected by extinction. The red leak is also more prominent for intrinsically red SEDs.
\end{enumerate}

Hence, we expect that airmass changes will cause shifts in the mean $u-v$ colour of a stellar population that gets stronger for redder populations, which means a colour-dependent stretch in our change index. From the axis definition it follows that changes in the $u$ band only cause anti-diagonal displacements. The strength of this displacement depends on airmass $A$, the SED and to a subtle level on dust extinction; it can be approximated as
\begin{equation}
  \Delta u = (1-A) (0.15 (AI_1+0.2)+0.049 E(B-V))   ~,
\end{equation}

where $AI_1$ is the age index at airmass 1. The centre panel of Figure~\ref{CI_3EFF} shows basic tracks of our change index from airmass 1 to 2, the range of SkyMapper observations, after applying the airmass correction. The SED dependence is not linear and causes an rms scatter of 0.003~mag at airmass 2, which is negligible relative to other uncertainties.

\subsubsection{Constructing insensitivity to redshift}\label{redshift}

For most galaxies in the nearby Universe redshifts are known independently and can be taken into account for the correct interpretation of the signals. We calculated synthetic colours for our grid of star-formation histories at redshifts 0 to 0.05 in steps of 0.01, in order to explore the consequences of moving the stellar spectra out of sync with the filter set. With increasing redshift we expect the decreasing alignment of the filters to turn the change index into an irrelevant age index, for which any blue-minus-red colour index could be used.

With redshift, the secular tracks move to the bottom right in the central diagnostic diagram, which means we need to recalibrate the zeropoint of the CI with a redshift compensation. We also see a change in the shape of the curves such that the scale changes, at which the age index needs to be added to remove age sensitivity. After fitting a quadratic redshift dependence to the resulting colours, we find the redshift-independent version of the indices as:
\begin{align}
  CI(z)&=CI-(2.8z+112z^2)\times AI_{v2} +4.8z+60z^2  & \nonumber \\
  AI(z)&=AI-7.22z+10z^2  ~, &
\end{align}
which we plot in Figure~\ref{CI_3EFF}.
As expected, the change index becomes ever more similar to an age index itself with increasing redshift, such that the age-insensitive version of the change index becomes less and less sensitive. At redshift $z=0.04$ the change-index diagnostic has halved in sensitivity, and above it gradually disappears.

\subsubsection{Effect of emission lines}\label{Elines}

The above spectral synthesis ignored nebular emission lines. This is mostly not a problem, as the equivalent width (EW) of emission lines is small relative to the width of the broad passbands. Star-forming galaxies have H$\alpha$ line widths ranging from 5 to 100~$\AA$ in extreme cases, with a median value of 15~$\AA$ \citep{KK83,Bloom17}; they lift the $r$ band flux by 1\% in the median and 6\% in exceptional cases.

However, the $u$ and $v$ bands underlying the change index are narrower. At the redshifts considered in this work, the main effect is the OII emission line entering the $v$ band. An OII emission line with 15~$\AA$ EW would lift the $v$ band flux already by 5\%, which is starting to be a relevant effect for the change index. \citet{BlantonLin00} measured the equivalent width of star-forming galaxies in the Las Campanas Redshift Survey (LCRS, at a median redshift of $\sim 0.1$) and found that less than 1\% of Milky Way-scale galaxies have O\textsc{ii} lines with EW $>$15~$\AA$. However, SMC-scale dwarf galaxies have lines stronger than this already in 5\% of cases. Also, Seyfert galaxies show significant circumnuclear OII ionisation regions \citep[e.g.][]{Dopita15}, posing an ambiguity between quenching regions and AGN Narrow-Line Regions (NLRs).

\begin{figure*}
\begin{center}
\includegraphics[angle=270,width=\columnwidth]{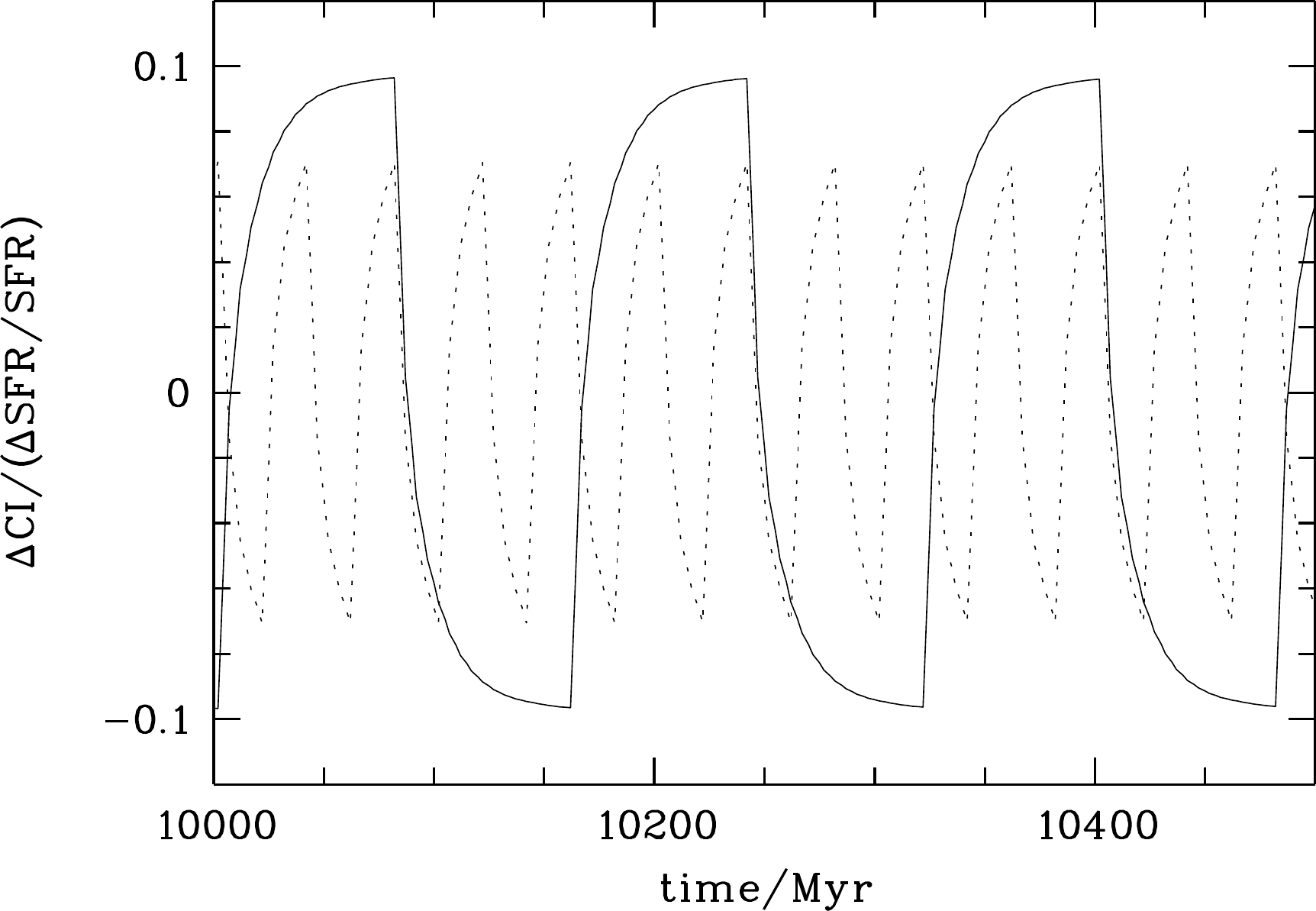}
\hspace{3mm}
\includegraphics[angle=270,width=\columnwidth]{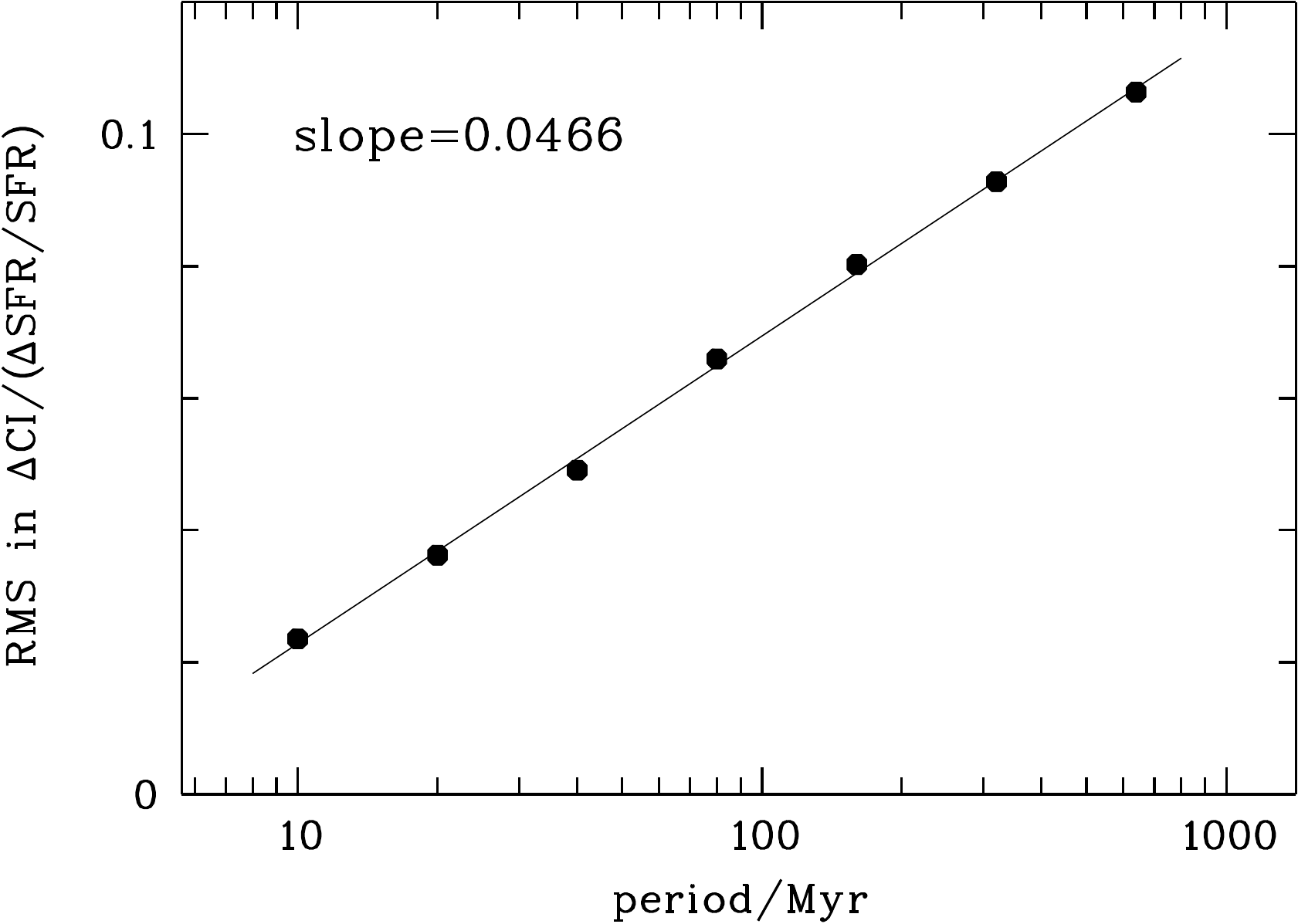}
\caption{{\it Left:} Response of the change index to comb-shaped waves in star-formation rate with wave periods of 40~Myr and 160~Myr. {\it Right:} Time-averaged RMS in the change index for wave periods from 10 to 640~Myr.}\label{CI_rms}
\end{center}
\end{figure*}

\subsubsection{Self-calibration and its limitations}

We also consider the possibility of unintended offsets in the zeropoint calibration of the images. Such offsets would shift the change index globally for the whole galaxy or indeed all galaxies at the same redshift, e.g. a group with satellites contained in the image. Every spatial resolution element of every galaxy in the group would thus appear to be quenched or bursting, irrespective of size, mass, surface brightness and overall SED. Such synchronous changes are physically unlikely, albeit not impossible. If a common offset is seen, the likely explanation will be calibration issues in the input data.

In this case, a self-calibration approach could remove the mean change across a galaxy or whole group. The residuals would be spatial changes in the change index that would at least reveal how some parts of a galaxy, or some satellites in a group, are being quenched faster than others. Obviously, however, the removal of a mean signal won't help if only one data point is available such as an unresolved single dwarf galaxy without further reference at the same redshift.

\subsection{Sensitivity to star formation stochasticity}

So far, we have modelled star formation histories as smooth functions, but in reality galaxies may experience short-term fluctuations in star formation, which will cause noise in the change index. This could be especially pronounced in small localised parts of a galaxy. In this section, we investigate the noise response of the change index to short-term noise in the SFR, using a Fourier-style approach. 

We begin with the smooth $\tau=10$~Gyr star formation history, and instead of applying a single burst or quenching event after 10~Gyr, we let the SFR oscillate up and down by $\pm 1\%$; a small amplitude helps to reveal the strength with which the variation propagates into the measured signal in the linear regime before the signal effect saturates. We choose a range of time scales, starting from 10~Myr duration for a full wave and realise waves as a comb with a 5~Myr top-hat bin of elevated SFR followed by a 5~Myr top-hat bin of reduced SFR. Longer waves are modelled as combs with wave period doubling at each step, up to more than 1~Gyr (see Figure~\ref{CI_rms}, left panel). 

We then measure the root-mean square of the change index in the comb section of the evolution, and find that the change index responds with nearly linear proportionality to the wave period up to periods of nearly 1~Gyr. This is expected as the change index behaves linear for small signals and deviations build up during any one-sided half-wave, only to be compensated during the counter-acting half-wave. For longer time periods, the comb becomes locally similar to a burst or quenching event, and the response looks just as expected from the results before.

The right panel of Figure~\ref{CI_rms} shows the resulting rms value in CI in mag per fractional change of SFR. Waves with a 80~Myr period, e.g., propagate into the CI with 1/15th of their amplitude. A 15\% wave amplitude in SFR would then lead to an rms variation in CI by 1\% (or $\sim 0.01$~mag). We thus conclude that stochastic variations in the SFR are not causing too much noise, while strong and persistent changes will be measured as intended.

\subsection{Sensitivity to active galactic nuclei (AGN)}

A large fraction of galaxies, especially at higher mass, are host to an active nucleus, even at redshift $z\approx 0$, where nuclear activity is subdued relative to higher redshift. We consider two regimes in the impact of nuclear activity on the spectrum:
\begin{enumerate}
\item Seyfert-2 and LINER galaxies have continuum spectra that look much like the pure stellar populations considered before, with the addition of narrow emission lines. The main impact of narrow emission lines is through the appearance of the OII line in the $v$ band, which gets more prominent as redshift increases and the line moves from the edge of the filter into its main transmission range. In Seyfert-2 galaxies, Extended Narrow-Line Regions (ENLRs), also known as ionisation cones, can cover kiloparsec-scales \citep{Schmitt03} and extend vertically out of the disk of a galaxy \citep[see examples from the S7 Survey in][]{Thomas17}. Where the OII emission exceeds 60~$\AA$ in equivalent width, the $u-v$ colour can appear as red as an instantaneous quenching event even when the star-formation rate is constant. This means that extremely strong ionisation regions could be discovered via extremely red $u-v$ colours. On the other hand, it will be difficult to disentangle ENLRs in Seyfert-2 galaxies from genuine quenching of star formation, when only SkyMapper colours are used.

\item In addition, Seyfert-1 galaxies have nuclear spectra, where an added quasar-like SED can dominate the light output. While Seyfert-1 galaxies are rare, their nuclear SEDs are entirely different from the stellar populations considered before. Seyfert-1 galaxies will exhibit cores that are somewhere on a mixing trajectory between a pure quasar SED and the host galaxy spectra studied before. The mixing ratio will not only depend on the intrinsic luminosity of the AGN and the surface brightness of the stellar core, but also on the physical resolution and hence the distance of the galaxy. Below we analyse the SkyMapper colours of a $z\approx 0$ quasar SED.
\end{enumerate}

We take the quasar spectral template from \citet{VandenBerk01}, separate the emission-line contour from the continuum and reassemble them into a grid of quasar properties. We span a 2D-space of quasar templates using a range of slopes multiplied onto the mean continuum (from $\Delta \alpha = +1$ to $-1$), and independently vary the intensity of the emission-line spectrum by factors of $1/3$ to $3\times$ relative to the mean. 

We calculate the SkyMapper colours for the grid and show the mean quasar colour as an asterisk in Figure~\ref{CC}; in $vgriz$ bands an average quasar-like SED cannot be separated from the stellar spectral tracks. However, the $u-v$ colour is significantly different, ranging from $-0.2$ to $+0.05$ for the whole grid, which is entirely outside the plot range in Figure~\ref{CC}. The clearest distinction is in the change index, where the entire quasar-like SED grid ranges from values of $-0.75$ to $-0.5$~mag, vastly exceeding the colour signal from the strongest starbursts.

In summary, nuclei dominated by Seyfert-1 light are easily recognised, while weak Seyfert-1 cores might not be distinguishable from nuclear starbursts with SkyMapper colours alone. Similarly, moderate NELGs in Seyfert-2 galaxies might not be distinguishable from quenching regions. A detailed exploration of the sensitivity limits for Seyfert detection is beyond the scope of this paper.

\subsection{Challenges posed by foreground stars}

Obviously, Galactic stars may populate the scenery around the line of sight to galaxies, and are often found projected in front of a galaxy's face. We thus ask how the presence of such stars may confuse the interpretation of the colour maps. Bright stars against faint galaxy background should stand out as small regions of constant colour, and could hence be deliberately ignored by masking part of the background galaxy from view. But if the galaxy brightness changes rapidly across the extend of a star's PSF, it can appear with a colour gradient in a colour map and may be mistaken for a galaxy feature.

Figure~\ref{CC} also compares the colours of individual stars to those of integrated stellar populations. In the $gri$ colour space, the star colours are virtually the same as those of galaxies, and also their $u-v$ colours have significant overlap (which is not shown to keep the figure clearer). However, we can easily differentiate single stars and stellar populations by using $g-i$ in combination with $u-g$ and $i-z$. In terms of the change index, the stellar locus lies above the strongest quenching curve in Figure~\ref{CI_SMSDSS} (not shown, and partly outside the axis range). Blue stars in particular produce high CI values, ranging e.g. from $+0.27$ to $+0.41$ for F0-5IV/V stars. This gives hope that individual stars can be recognised as such and won't be misleading the interpretation of galaxies.

Still, when stars are faint and appear completely blended into the galaxy colour map, they can add noise to the colour maps and might make the interpretation harder. A systematic investigation of the practical impact is beyond the scope of this paper.

\section{FIRST RESULTS FROM SKYMAPPER}\label{data}

We explore the practical application of our star formation change index using Data Release 2 (DR2) of the SkyMapper Southern Survey \citep{Onken19}. DR2 is the first of the SkyMapper data releases that includes exposures from the SkyMapper {\it Main Survey}, which are exposed for 100~sec each and are read-noise limited only in the $u$ and $v$ bands. Given the typical surface brightness of nearby galaxies we need Main Survey images in $uvgi$ for the work in this paper. In DR2 these are available on one third of the Southern hemisphere.

\subsection{Matching images and background subtraction}

Colour indices are ratios of flux measurements, which get biased by offsets in the zero level of the flux scale; in the case of extreme flux ratios an offset-induced bias can diverge to infinity. Hence, accurate background subtraction is the most critical issue for the creation of colour index maps of a galaxy. Uncertainties in the background propagate directly into uncertainties of the colour index map. Colour indices of the highest surface-brightness portions in a galaxy, usually its core, will be limited by the relative zeropoint accuracy of the image pair. In DR2 relative zeropoints between passbands are mostly accurate to within 2\%. Regions of lower surface brightness will be increasingly biased by background offsets with decreasing flux. 

We apply several processing steps to minimise ratio artefacts from mismatching image properties. First, we subtract backgrounds from the images. Next we determine the mean PSF on each CCD frame and resample images to a common coordinate frame using PSFEx and SWarp\footnote{from the AstrOmatic toolbox at http://www.astromatic.net/}, including a PSF homogenisation that convolves all images to a common PSF with a 5~arcsec FWHM. In this paper, we ignore PSF variations across a CCD and use one fixed convolution kernel per CCD. This will introduce subtle position-dependent mismatches between the inputs of the colour maps. Finally, we scale images to common zeropoints and average them while ignoring pixel values that are masked as bad. 

We use a mesh grid-based background estimation method, similar to the one used by SExtractor \citep{BA96}, but with several improvements. We first build a source mask for each individual CCD image. With a series of binary morphological operations, such as erosion, dilation and opening, we estimate roughly which pixels are affected by sources. Because the seeing in our images varies a lot, the mask is designed to be relatively conservative, i.e. usually significantly larger than the isophotal area. This source mask is combined with the bad pixel mask to get a pure sky region.

\begin{figure}
\begin{center}
\includegraphics[width=\columnwidth]{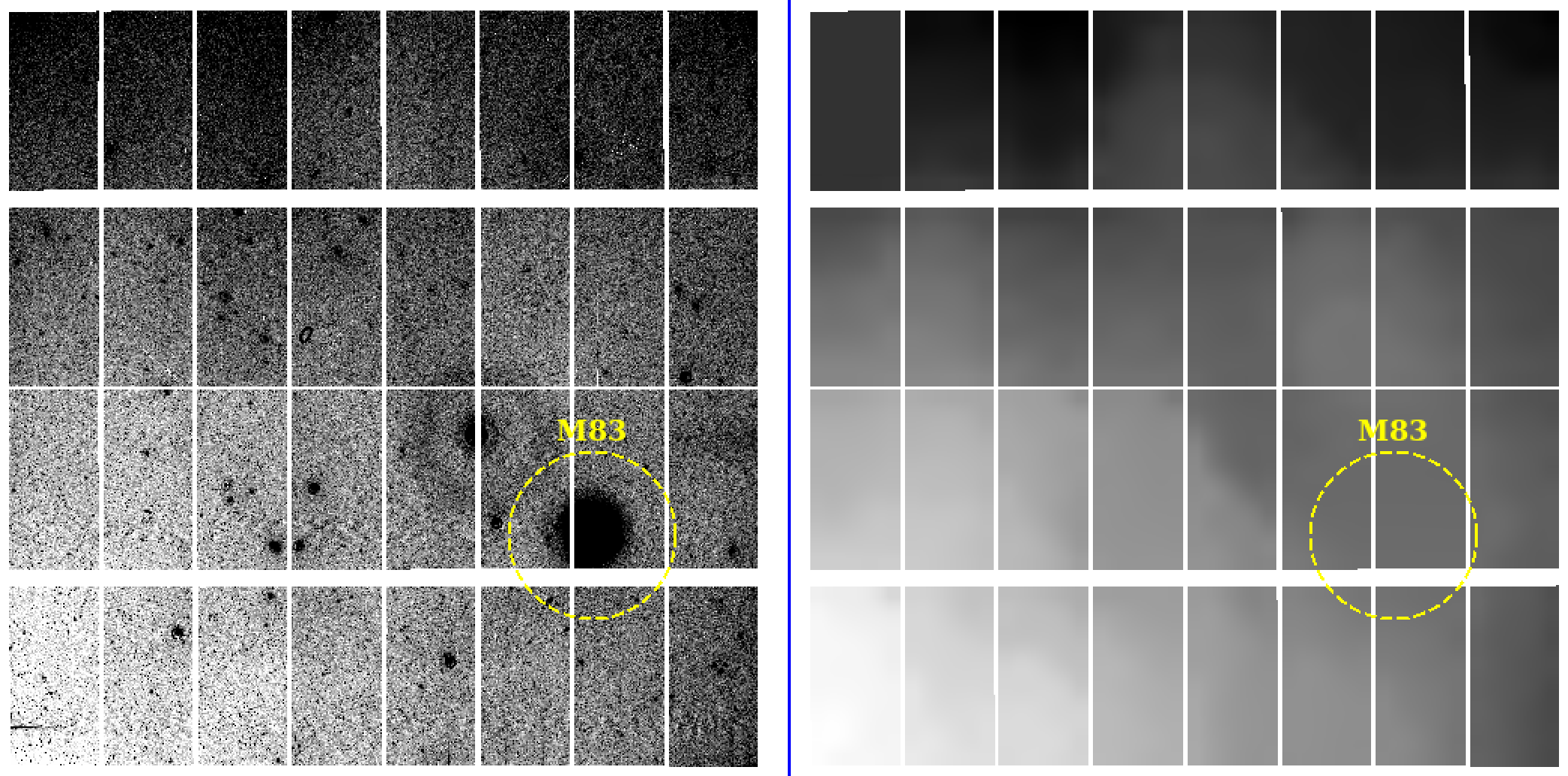}
\caption{{\it Left:} Co-added image in the region of the galaxy M 83. {\it Right:} Background fitted to the image; note that the galaxy does not appear to have biased the background map.}\label{background}
\end{center}
\end{figure}

\begin{figure}
\begin{center}
\includegraphics[width=\columnwidth]{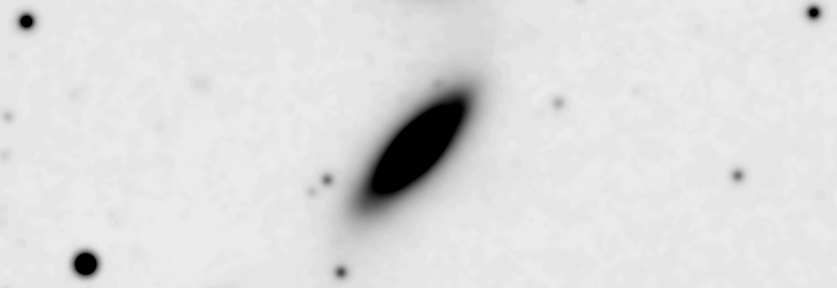}
\includegraphics[width=\columnwidth]{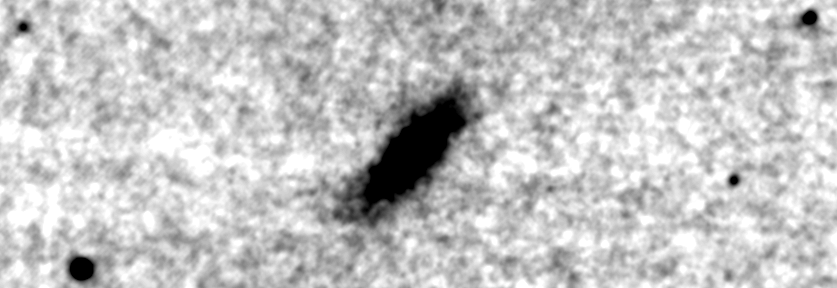}
\caption{High-contrast image of the galaxy NGC 1320, scaled to represent count rates from $-0.5$\% (white) to $+5$\% (black) of the central peak surface brightness of the galaxy. {\it Top:} $g$ band image. {\it Bottom:} $u$ band image showing a mildly lowered background level in the readout rows that are affected by the galaxy.}\label{background_hicon}
\end{center}
\end{figure}

We use a mesh size of $512 \times 512$ pixels, in order to capture the large scale background gradient and avoid contamination from large objects. Here, we assume that the true background does not have significant structure below the scale of $\sim 4$ arcmin (i.e. 512 pixels). In each mesh box, we use the unmasked pixels to calculate the background: we first do a 3.5 sigma-clipping to further remove any outliers not yet rejected by our mask. We then calculate the mode value ($2.5\times$ median $-1.5\times$ mean) of the remaining sky pixels as the background value for this mesh box.

We then combine the mesh boxes from all 32 CCD images in the same exposure to make a large-area map of the whole background in an exposure, with a spatial resolution of the mesh box size (similar to the SExtractor mini-background). Here, we assume the flatfield and bias correction are good enough to allow direct extrapolation between different CCD images. We note that our algorithm is robust against mild discontinuity between neighboring CCDs (see Figure~\ref{background}).

If too many pixels (>80\%) are masked in one mesh box, the box will be flagged as bad. The 80\% threshold is an empirical value, given that our source mask is relatively conservative and our mesh size is relative large. The bad mesh box implies a potential contamination from a large object or over-crowded regions. In order to preserve the outskirts of extended emission from large objects, such as a nearby galaxy, we also check the background values in boxes around the bad ones. We spread the bad flag to those neighbour boxes that have significantly higher background values ($>2.5\sigma$) than the overall background in the exposure. We also apply a median filter to further smooth the large background map, similar to what SExtractor does.

Finally, we fill the bad boxes with linear interpolation from good boxes with scipy.interpolate.griddata, which gives a reasonable approximation of the background below large objects or in over-crowded regions. We use the scipy.interpolate.RectBivariateSpline interpolation function to convert the low-resolution background map to one with the pixel resolution of the original image.

Generally speaking, our background estimation is robust against large nearby galaxies, without any prior knowledge of where these are. In Figure~\ref{background}, we show a direct comparison between the original image and the estimated background of one exposure near Messier 83. It is clear that the background below M83 does not show any signature of over-subtraction. However, if the whole field of view is dominated by objects, such as in the Galactic plane, we cannot find the true background without assistance from external data.

\begin{table*}
\caption{Example galaxies discussed in this paper. Morphological types are from 2MRS \citep{Huchra12} as are redshifts, except that of SDSSJ2114-0032 is from \citet{Pracy13}. $L_K$ is in units of ${\rm L}_\odot$. Object ID is from SkyMapper DR2.}
\label{sample} 
\centering          
\begin{tabular}{lrccccrlc}
\hline\hline       
Name 		& object ID & type & RA & Dec 		& $z$ 	& $\log L_K$ & Type & Figure \\ 
\hline
IC 976      & 110453626 & E+A    & 14:08:43.29&$-$01:09:41.3 & 0.005 &  9.85 &  S0/a & \ref{gal_images_EpA}\\
SDSSJ2114-0032 & 15795392 & E+A  & 21:14:00.54&$+$00:32:06.3 & 0.027 &  ?.?? &       & \ref{gal_images_EpA}\\
\hline
Mrk 1044    &  15344896 & Sy-1   & 02:30:05.52&$-$08:59:53.1 & 0.016 & 10.80 &  S0  & \ref{gal_images}\\
IC 1524     &   8637642 & Sy-1   & 23:59:10.72&$-$04:07:37.1 & 0.019 & 11.16 &  Sc  & \ref{gal_images}\\
\hline
NGC 7070    & 488344008 & galaxy & 21:30:25.33&$-$43:05:13.4 & 0.008 & 10.37 &  Scd  & \ref{gal_images2}\\
NGC 1320 (Mrk 607) &  21671048 & Sy-2   & 03:24:48.74&$-$03:02:32.1 & 0.009 & 10.79 & S0   & \ref{gal_images2}\\
\hline
MCG+00-04-112& 14034101 & Sy-2   & 01:23:21.30&$-$01:58:35.5 & 0.016 & 10.63 &  Sb   & \ref{gal_images3}\\
IC 1368     &  15829295 & Sy-2   & 21:14:12.59&$+$02:10:40.5 & 0.013 & 10.75 &  Sc   & \ref{gal_images3}\\
\hline
NGC 1321    &  21671053 & merger & 03:24:48.58&$-$03:00:56.0 & 0.009 & 10.79 &  S0/a & \ref{gal_images4}\\
%
\hline                  
\end{tabular}
\end{table*}

\begin{table*}
\caption{Core photometry of example galaxies in AB magnitudes. E+A galaxies have the reddest change index (CI), while Seyfert-1 galaxies have the bluest ones at levels not expected even for starbursts. The bright halo star projected against the galaxy NGC 1321 combines a blue $g-i$ colour with a red CI, which is not expected in any stellar population. Colour indices are dereddened using $E(B-V)$ values are from \citet{SFD98} and a \citet{F99} reddening law with $R_V=3.1$. Airmass values of all included Main Survey images range from 1.10 to 1.32.}
\label{sample2} 
\centering          
\begin{tabular}{lrcccrrrr}
\hline\hline       
Name 		 & object ID & type & $\mu_{u,\rm max}$ & $E(B-V)$ & CI $=$CI$_0$ & $(u-v)_0$ & $(v-g)_0$ & $(g-i)_0$ \\ 
\hline
IC 976      & 110453626 & E+A    & 21.95 & 0.054 & 0.25  & 0.52  & 1.04 & 0.53 \\
SDSSJ2114-0032 & 15795392 & E+A  & 23.16 & 0.133 & 0.25  & 0.51  & 1.19 & 0.56 \\
\hline
Mrk 1044    &  15344896 & Sy-1   & 18.61 & 0.034 &$-0.46$&$-$0.10& 0.27 & 0.30 \\
IC 1524     &   8637642 & Sy-1   & 20.32 & 0.037 &$-0.30$& 0.09  & 0.94 & 0.81 \\
\hline
NGC 7070    & 488344008 & galaxy & 21.67 & 0.030 & 0.05  & 0.37  & 1.02 & 0.65 \\
NGC 1320 (Mrk 607)&21671048& Sy-2& 22.31 & 0.050 & 0.05  & 0.38  & 1.51 & 0.98 \\
\hline
MCG+00-04-112& 14034101 & Sy-2   & 22.49 & 0.046 & 0.16  & 0.54  & 1.33 & 0.97 \\
IC 1368     &  15829295 & Sy-2   & 22.25 & 0.076 & 0.13  & 0.49  & 1.46 & 1.04 \\
\hline
NGC 1321    &  21671053 & merger & 21.70 & 0.050 & 0.16  & 0.44  & 1.37 & 0.73 \\
companion   &  21671052 & halo star&21.08& 0.050 & 0.16  & 0.41 & 0.85 & 0.40 \\
\hline                  
\end{tabular}
\end{table*}

\begin{figure*}
\begin{center}
\includegraphics[width=0.19\textwidth,frame]{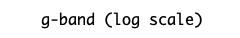}
\includegraphics[width=0.19\textwidth,frame]{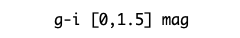}
\includegraphics[width=0.19\textwidth,frame]{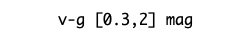}
\includegraphics[width=0.19\textwidth,frame]{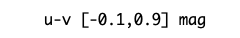}
\includegraphics[width=0.19\textwidth,frame]{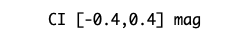}
\includegraphics[width=0.19\textwidth,frame]{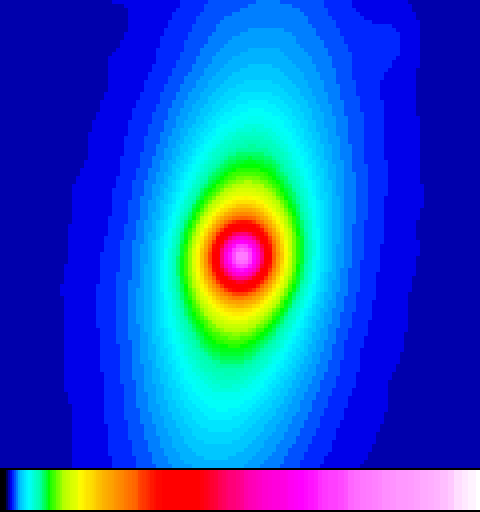}
\includegraphics[width=0.19\textwidth,frame]{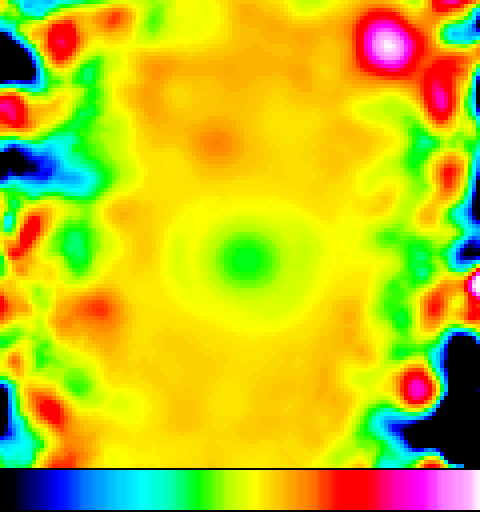}
\includegraphics[width=0.19\textwidth,frame]{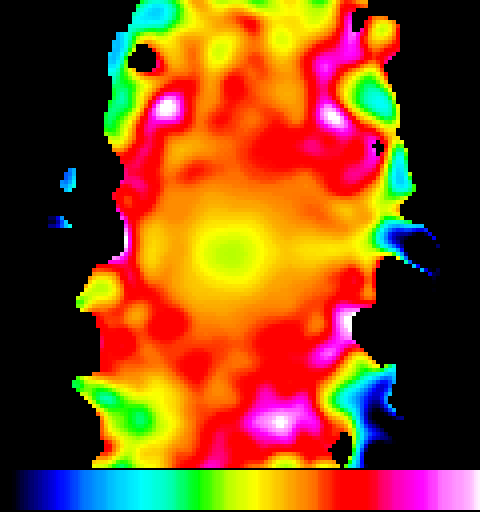}
\includegraphics[width=0.19\textwidth,frame]{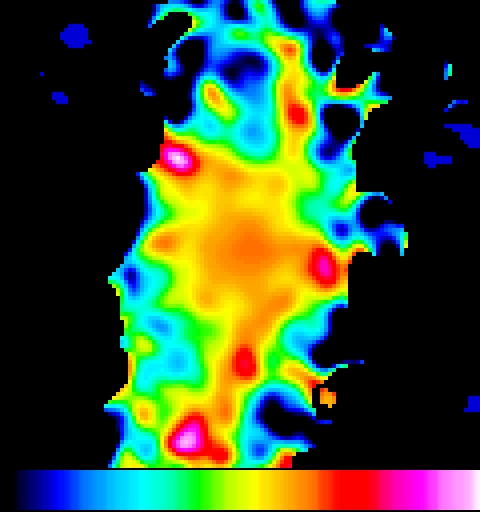}
\includegraphics[width=0.19\textwidth,frame]{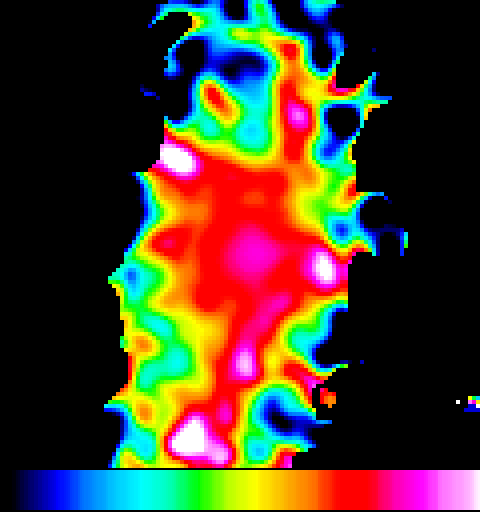}
\includegraphics[width=0.19\textwidth,frame]{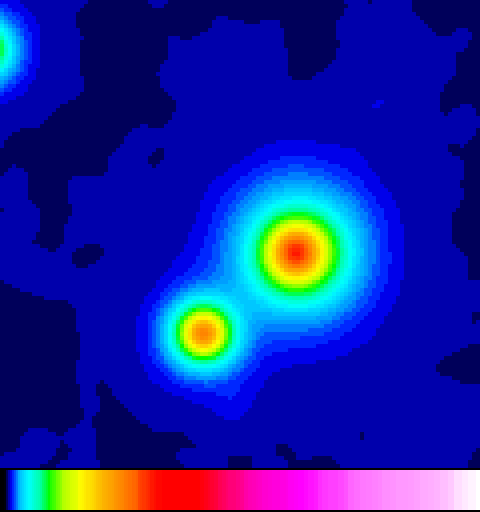}
\includegraphics[width=0.19\textwidth,frame]{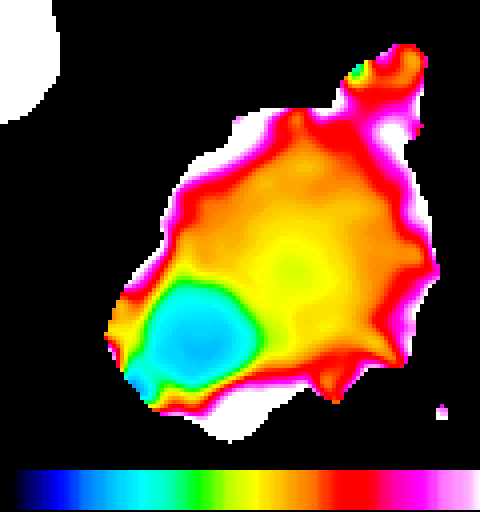}
\includegraphics[width=0.19\textwidth,frame]{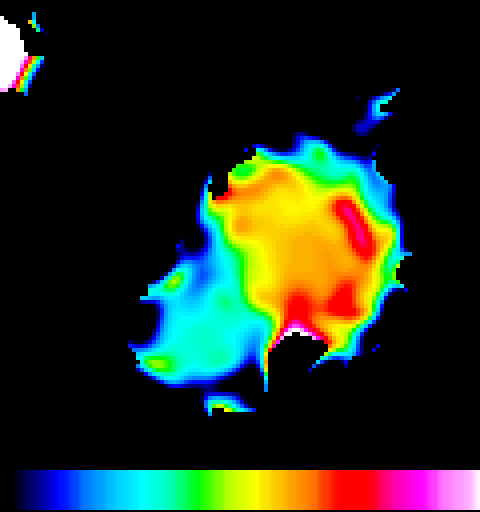}
\includegraphics[width=0.19\textwidth,frame]{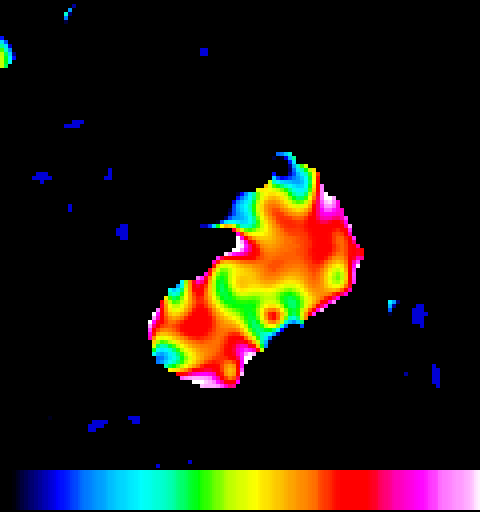}
\includegraphics[width=0.19\textwidth,frame]{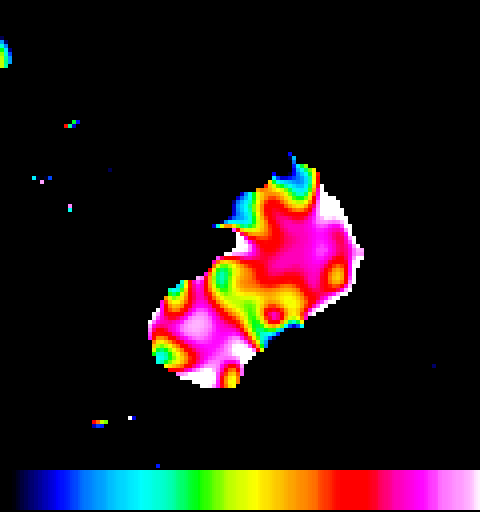}
\caption{Known E+A galaxies, from left to right: 
$g$ band image (white level = 25~mag arcsec$^{-2}$), followed by colour maps in $g-i$, $v-g$, $u-v$ 
and change index map (size: $1\arcmin \times 1\arcmin $). Black-to-white ranges are indicated in the header row.
Top: IC 976 at $z=0.005$. Bottom: SDSS J2114$-$0032 at $z=0.027$. 
Both galaxy cores are blue in $g-i$ and red in $u-v$ and the change index.
}\label{gal_images_EpA}
\end{center}
\end{figure*}

\subsection{Background over-subtraction in DR2}

Despite our careful background subtraction, DR2 images of large extended sources are affected by an over-subtraction of the bias, which results from the bias-removal procedure: the CCD readout electronics of SkyMapper are temporally unstable on timescales that are shorter than the read-out time of a single CCD row. Hence, the CCD overscan contains no information on the true bias level and its variation across a given row of any image. As described by \citet{Wolf18} the shape of the bias across any image row is determined by an empirical PCA-based process after masking detectable objects from a frame and removing a large-scale background. 

As part of this procedure, outer regions of a galaxy at lower surface brightness will not be masked, but instead be mistaken as bias, and will be partially subtracted from the image. While the procedure assumes a limit for the bias amplitude, this affects particularly the $u$ and $v$ band images, which are always read-noise limited except in the cores of bright galaxies and in most point sources. The effect is that colour index maps in $u-v$ based on DR2 can only be trusted in the inner parts of galaxies, or on point sources.

In DR2, the over-subtraction of background under large galaxies is strongest in the $u$ band and can be as large as 10~counts. The galaxies discussed below have central surface brightness levels of $\mu_{u,AB}$ from 22.5 to 21.7~mag~arcsec$^{-2}$ after convolution to a $5\arcsec$ PSF (except for the objects with bright blue cores). A ten-count bias would redden their $u-v$ colour by 0.06 to 0.1~mag in their cores and by more in their outskirts. As we will see below, a consequence of this is that in most $u-v$ colour maps the outermost galaxy contours seem to have their colours diverging to the red beyond physically plausible values, irrespective of physically meaningful colour gradients in the brighter parts of the galaxies.

\subsection{Noise levels in DR2}

In the SkyMapper Main Survey all images are exposed for 100~sec, but zeropoints vary strongly among the bandpasses. Those with the lowest sensitivity by far are the relatively narrow, short-wavelength $u$ and $v$ bands, and their errors will dominate any error consideration around the change index.

In average exposures the noise level in $u$ and $v$ is 10~counts per pixel and results mostly from read-out noise. Together with median image zeropoints, we find that at a surface brightness of $\mu_{u,AB} \approx 22$~mag~arcsec$^{-2}$ we get an rms error of 0.1~mag in the $u-v$ colour in a cell of 1~arcsec$^2$. After smoothing over a box of $5\arcsec \times 5\arcsec$ this becomes a tolerable 2\% (0.02~mag) error. DR2 has usually three exposures per band, which we have combined here into deeper images. Hence, we get a 2\% error at 22.6~mag~arcsec$^{-2}$, which is satisfied by every galaxy discussed in this paper except one.

Below, we discuss several example galaxies and their colour and change index maps. In those, we choose to black out regions where we expect the colour index to have a $1\sigma$-error greater than 0.5~mag. We simply estimate a flux threshold for the noisiest band in a colour index from the global noise properties of the input image and drop all regions with flux below the threshold.

\begin{figure*}
\begin{center}
\includegraphics[angle=270,width=0.49\textwidth]{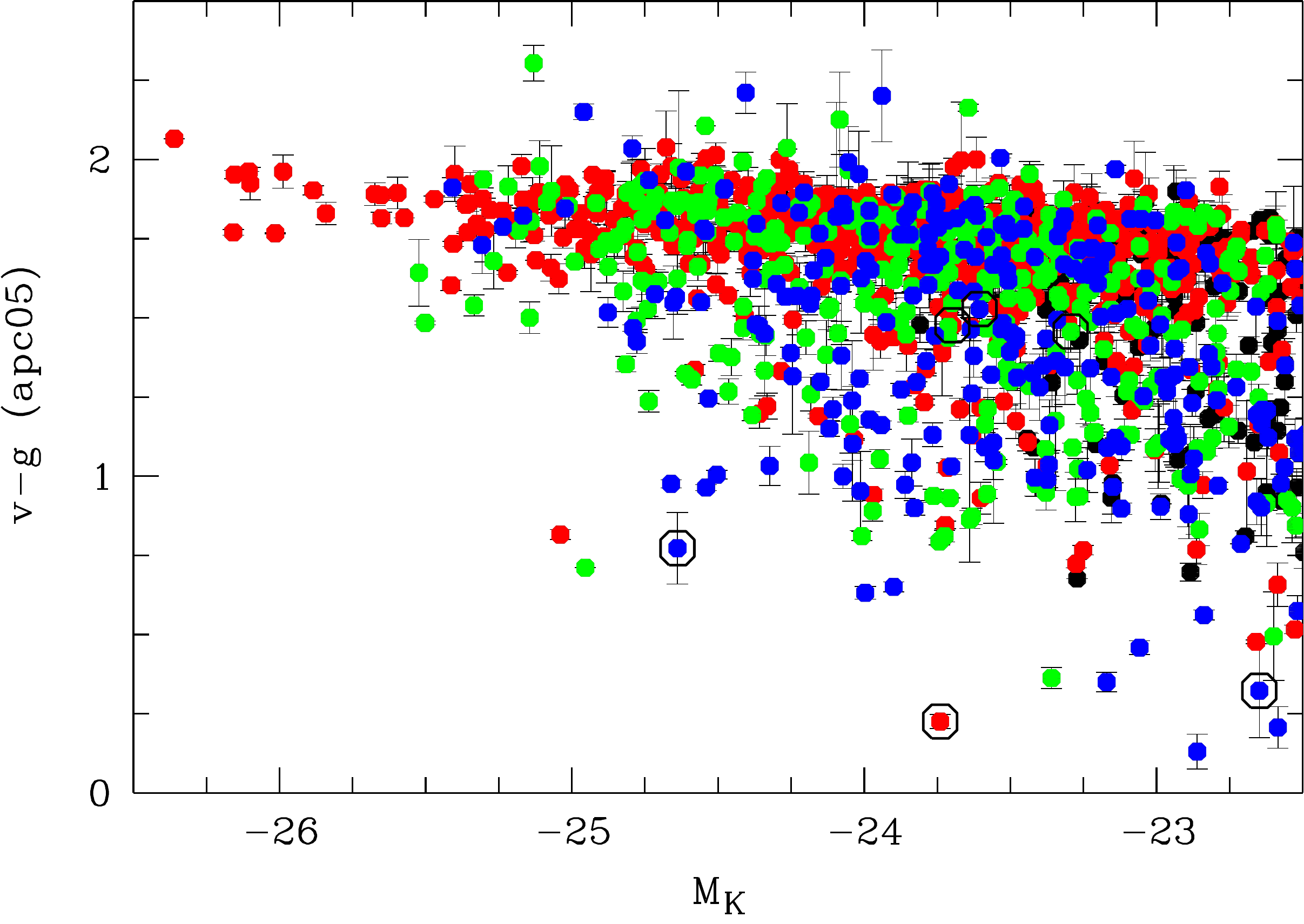}
\hfill
\includegraphics[angle=270,width=0.49\textwidth]{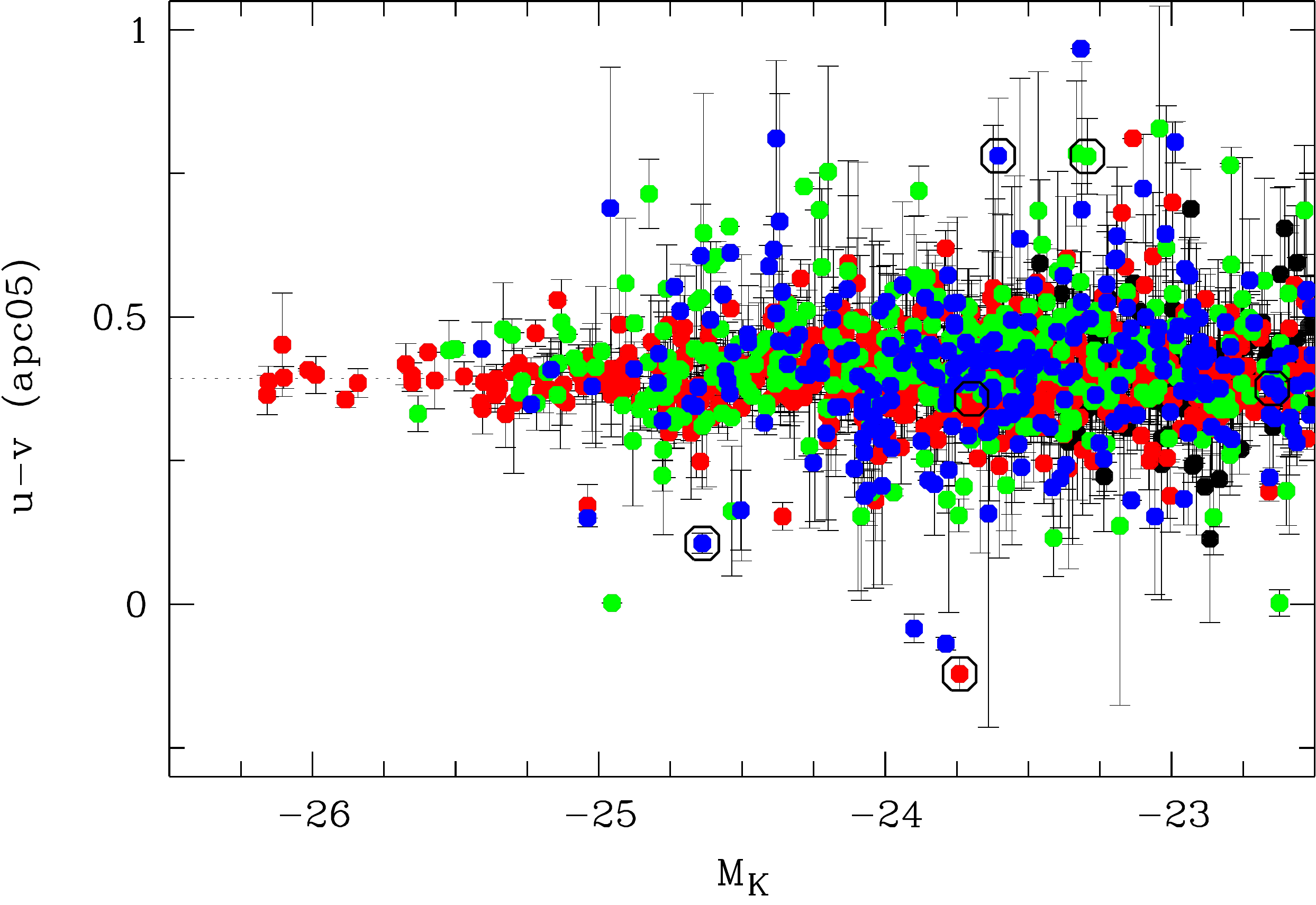}
\caption{Core colour-vs.-total magnitude diagrams of 2MRS galaxies based on $5\arcsec$-aperture photometry. Shown are $\sim$1,300 galaxies with complete Main Survey data at $M_K<-22.5$, $E(B-V)_{\rm SFD}<0.1$ and $v=[1000,6000]$~km s$^{-1}$, colour-coded by 2MASS morphology (red: E/S0, green: Sa-Sc, blue: Sd-Irr, black: unclassified). The six encircled objects are discussed in Figures~\ref{gal_images} to \ref{gal_images3}. 
}\label{CMDs}
\end{center}
\end{figure*}

\subsection{Known nearby E+A galaxies}

We expect the SkyMapper $u$ and $v$ filters and the change index to be primarily useful as a quenching indicator. Hence, we first look at known E+A galaxies at $z<0.03$, where we expect useful results. We start from the E+A galaxy samples by \citet{Blake04} and \citet{Pracy12,Pracy13,Pracy14}, which contain eight Southern objects at $z<0.03$, but only two objects with Main Survey coverage in DR2 (see Tables~\ref{sample}, \ref{sample2} and Figure~\ref{gal_images_EpA}): 

The first is the S0 galaxy IC~976 at $z=0.005$, which is listed as TGN274Z137 in \citet{Blake04} and called E+A6 in the spectroscopic study by \citet{Pracy12}; given a distance of only $\sim 20$~Mpc, IC~976 has a spatial extent of several arcminutes and provides great spatial resolution. Figure~\ref{gal_images_EpA} shows this galaxy in the top row. Both the $g-i$ and the $v-g$ colour map reveal strong blue cores resulting from young stars. In contrast, the $u-v$ colour and change index maps show pronounced red cores and clarify that star formation in the core has been quenched; the youngest stars providing the blue $vgi$ colours are A stars, as expected in E+A populations, while OB stars are absent causing the red $u-v$ colour. A galaxy with continuing star formation in the core would show a neutral $u-v$ colour and a nuclear starburst would lead to a blue core in $u-v$.

The second galaxy is SDSSJ2114$-$0032 at $z=0.027$ and has been studied in detail by \citet{Pracy13}. Owing to the larger distance of $\approx 115$~Mpc and its smaller intrinsic size, this galaxy appears much smaller than E+A6 (see second row in Figure~\ref{gal_images_EpA}). While the image contains an F-type halo star just 15 arcsec to the lower left of the galaxy core, the $g-i$ colour map of the galaxy still shows a clear colour gradient towards a blue core. The $v-g$ colour map is less clear and perhaps affected by the star or by the moderately high redshift relative to the expected working range of the change index. At this redshift the Balmer break has shifted by 10~nm into the relatively narrow $v$ band, lowering its flux in comparison to both the $u$ and the $g$ band. In the $u-v$ and change index maps we see again a red core reminiscent of a recently quenched population (as expected, the single F-type star is nearly as red in $u-v$). 

Note, that according to \citet{SFD98} the second galaxy is reddened by Milky Way dust with $E(B-V)_{\rm SFD}=0.133$, which corresponds to $(E(u-v),E(v-g),E(g-i)) \approx (0.04,0.14,0.19)$ assuming a \citet{F99} reddening law with $R_V=3.1$. The colour maps in this paper are as observed and not corrected for reddening. However, reddening is subtracted out in the colour values of Table~\ref{sample2}, and the change index map is by design insensitive to reddening with $R_V=3.1$. The shown change index maps are not corrected for redshift effects or for airmass effects that change the $u$ band transmission curve.

Both these galaxies had been observed with the integral-field Wide Field Spectrograph \citep[WiFeS,][]{Dopita10} at the ANU 2.3m-telescope at Siding Spring Observatory. \citet{Pracy12} and \citet{Pracy13} report their spectra and show strong signs of a dominant A~star population in their cores. 

\begin{figure*}
\begin{center}
\includegraphics[width=0.19\textwidth,frame]{title_g.png}
\includegraphics[width=0.19\textwidth,frame]{title_gmi.png}
\includegraphics[width=0.19\textwidth,frame]{title_vmg.png}
\includegraphics[width=0.19\textwidth,frame]{title_umv.png}
\includegraphics[width=0.19\textwidth,frame]{title_ci.png}
\includegraphics[width=0.19\textwidth,frame]{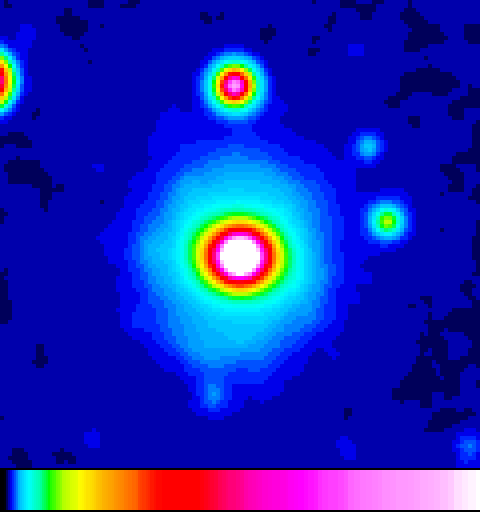}
\includegraphics[width=0.19\textwidth,frame]{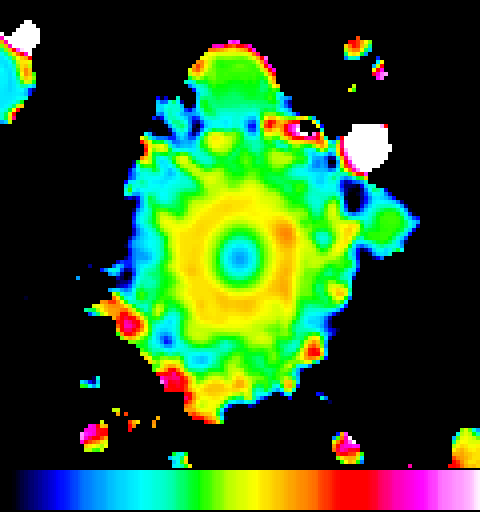}
\includegraphics[width=0.19\textwidth,frame]{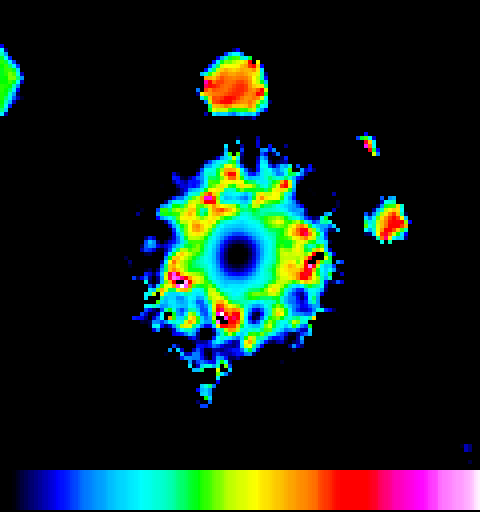}
\includegraphics[width=0.19\textwidth,frame]{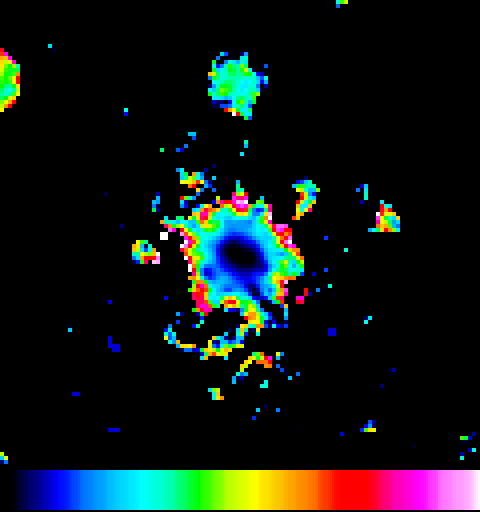}
\includegraphics[width=0.19\textwidth,frame]{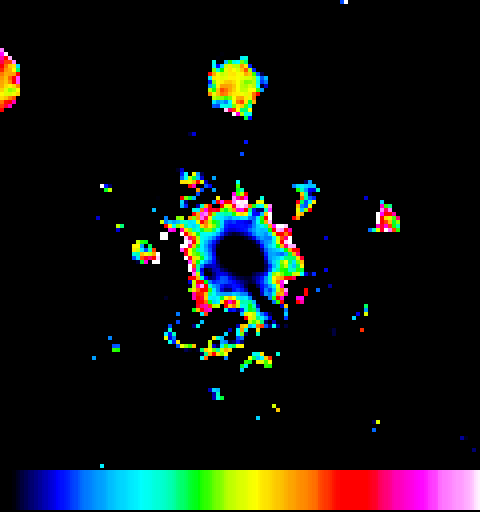}
\includegraphics[width=0.19\textwidth,frame]{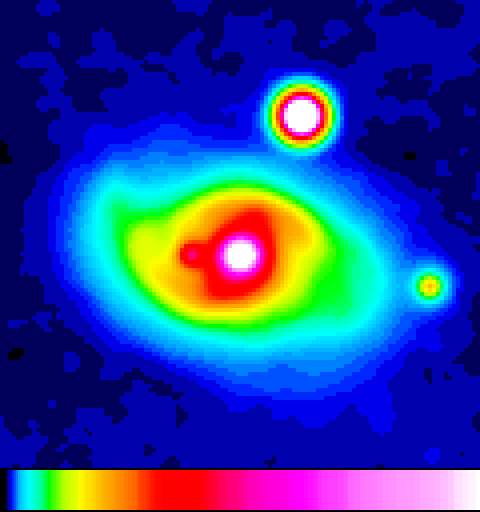}
\includegraphics[width=0.19\textwidth,frame]{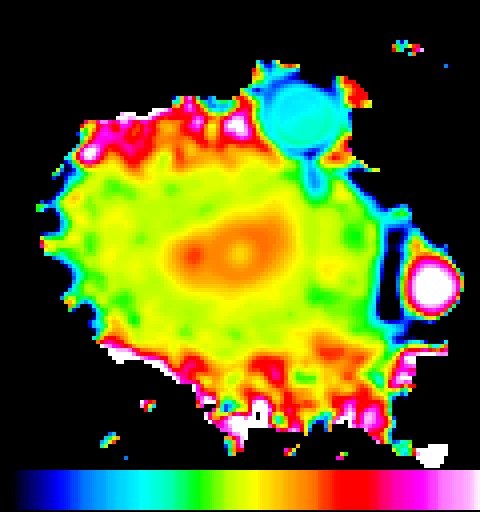}
\includegraphics[width=0.19\textwidth,frame]{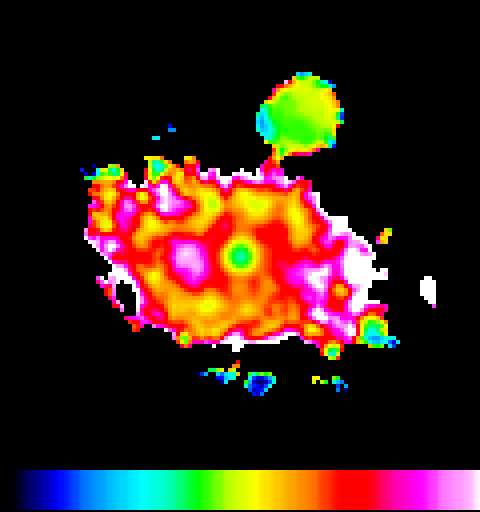}
\includegraphics[width=0.19\textwidth,frame]{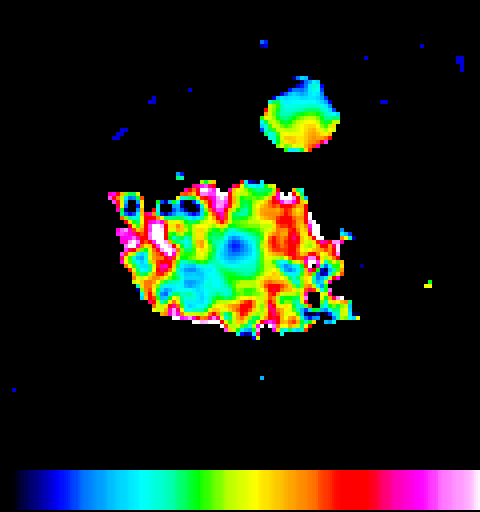}
\includegraphics[width=0.19\textwidth,frame]{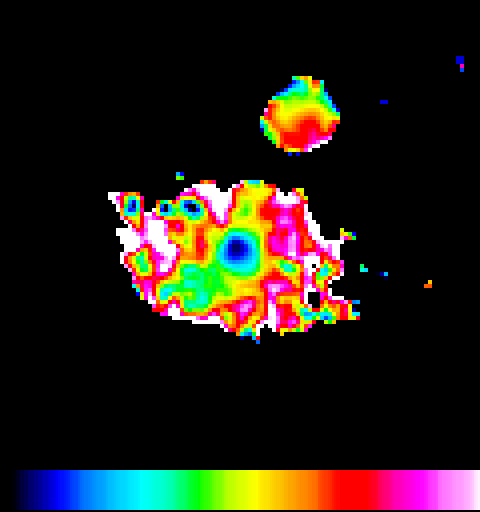}
\caption{Galaxies with blue cores in $u-v$ (see Figure~\ref{gal_images_EpA} for details). 
Top: Sy-1 galaxy Mrk 1044. Bottom: Sy-1 galaxy IC 1524. The extremely blue $u-v$ and change index colours are inconsistent with a starburst and can only be explained with a type-1 AGN.
}\label{gal_images}
\end{center}
\end{figure*}

\begin{figure*}
\begin{center}
\includegraphics[width=0.19\textwidth,frame]{title_g.png}
\includegraphics[width=0.19\textwidth,frame]{title_gmi.png}
\includegraphics[width=0.19\textwidth,frame]{title_vmg.png}
\includegraphics[width=0.19\textwidth,frame]{title_umv.png}
\includegraphics[width=0.19\textwidth,frame]{title_ci.png}
\includegraphics[width=0.19\textwidth,frame]{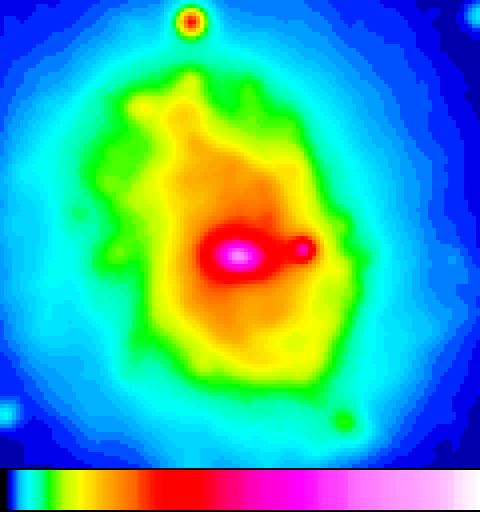}
\includegraphics[width=0.19\textwidth,frame]{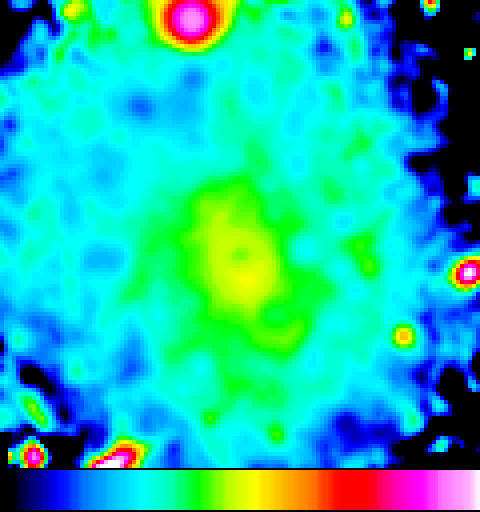}
\includegraphics[width=0.19\textwidth,frame]{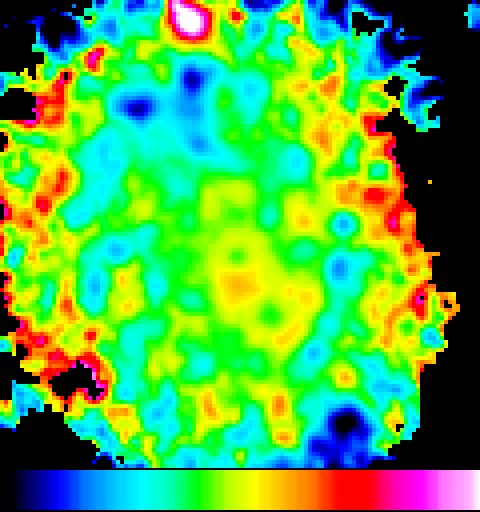}
\includegraphics[width=0.19\textwidth,frame]{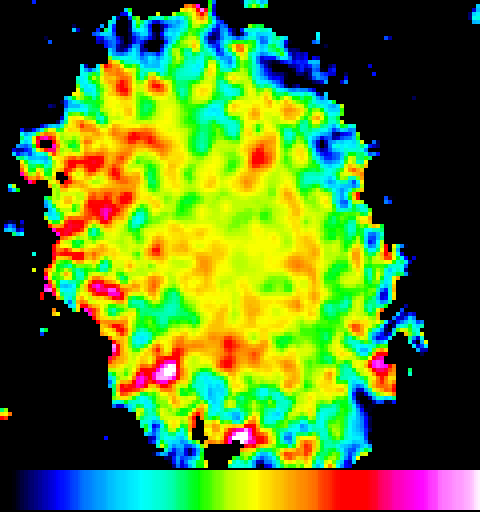}
\includegraphics[width=0.19\textwidth,frame]{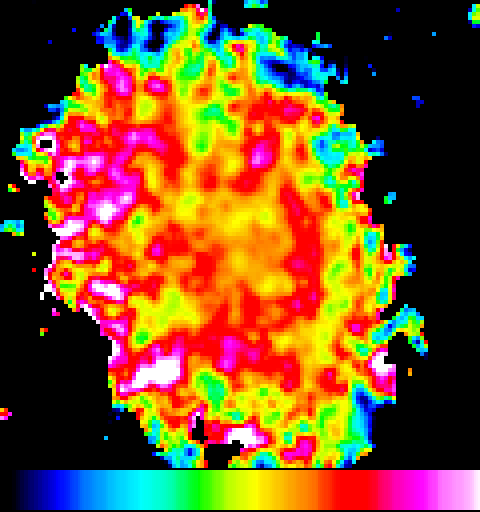}
\includegraphics[width=0.19\textwidth,frame]{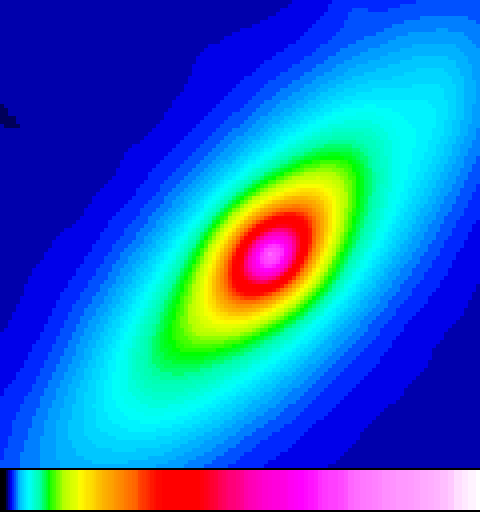}
\includegraphics[width=0.19\textwidth,frame]{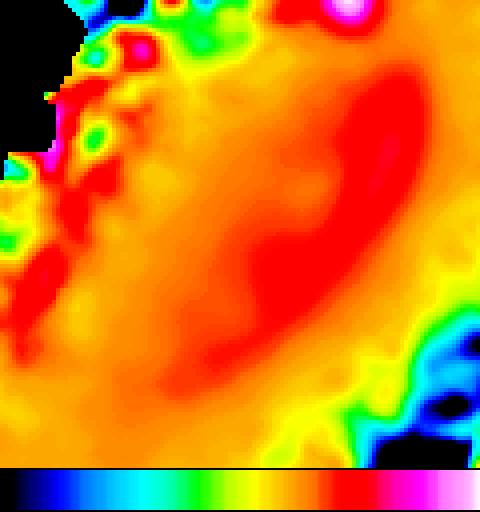}
\includegraphics[width=0.19\textwidth,frame]{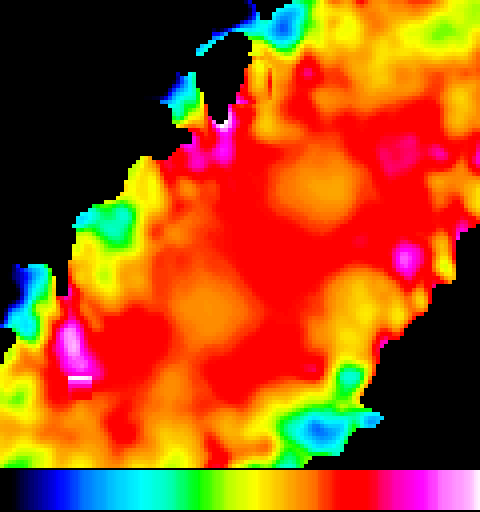}
\includegraphics[width=0.19\textwidth,frame]{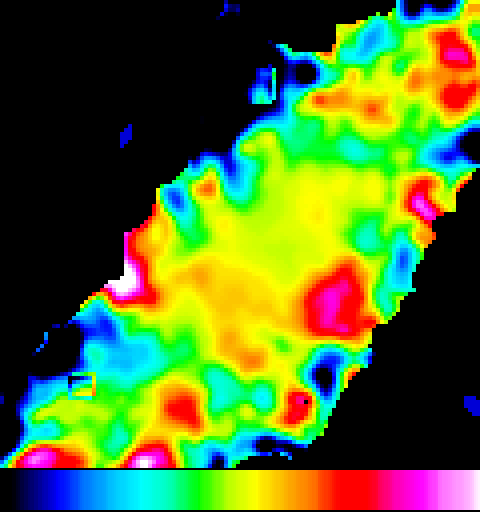}
\includegraphics[width=0.19\textwidth,frame]{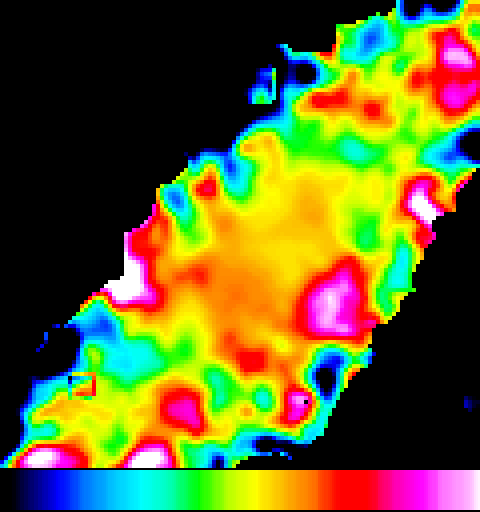}
\caption{Galaxies with intermediate cores in $u-v$ (see Figure~\ref{gal_images_EpA} for details).
Top: galaxy NGC 7070. Bottom: Sy-2 galaxy NGC1320. Note that the dark red features in the $g-i$ image of NGC 1320 are coincident with dark dust lanes seen in HST images of this galaxy.
}\label{gal_images2}
\end{center}
\end{figure*}

\begin{figure*}
\begin{center}
\includegraphics[width=0.19\textwidth,frame]{title_g.png}
\includegraphics[width=0.19\textwidth,frame]{title_gmi.png}
\includegraphics[width=0.19\textwidth,frame]{title_vmg.png}
\includegraphics[width=0.19\textwidth,frame]{title_umv.png}
\includegraphics[width=0.19\textwidth,frame]{title_ci.png}
\includegraphics[width=0.19\textwidth,frame]{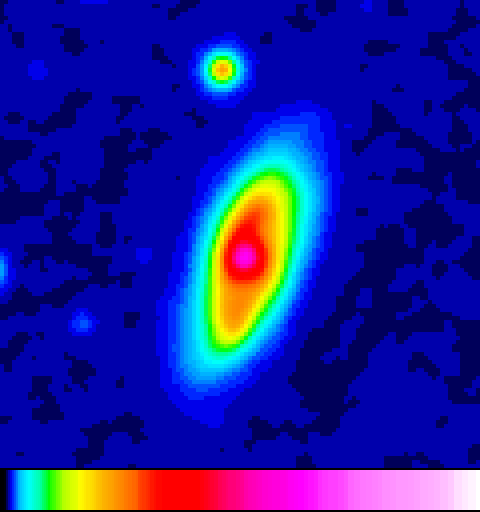}
\includegraphics[width=0.19\textwidth,frame]{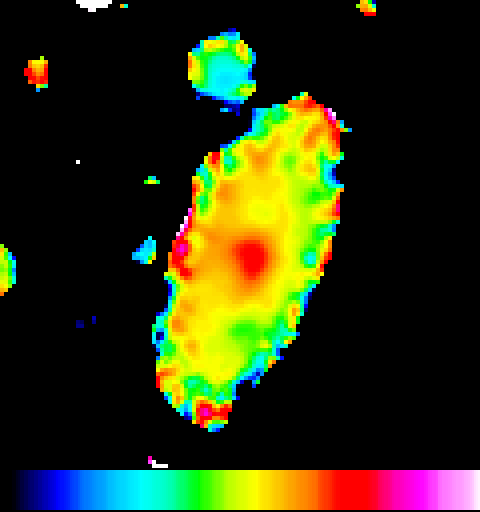}
\includegraphics[width=0.19\textwidth,frame]{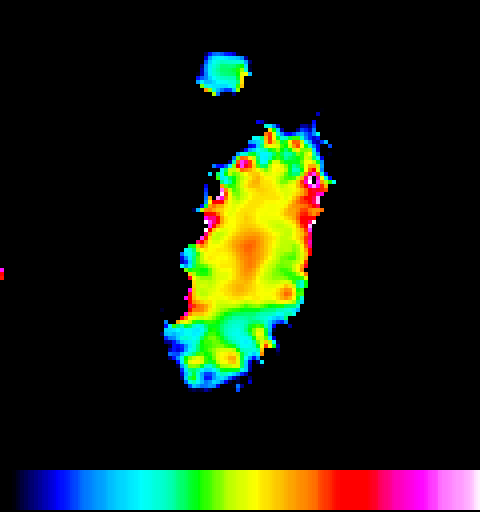}
\includegraphics[width=0.19\textwidth,frame]{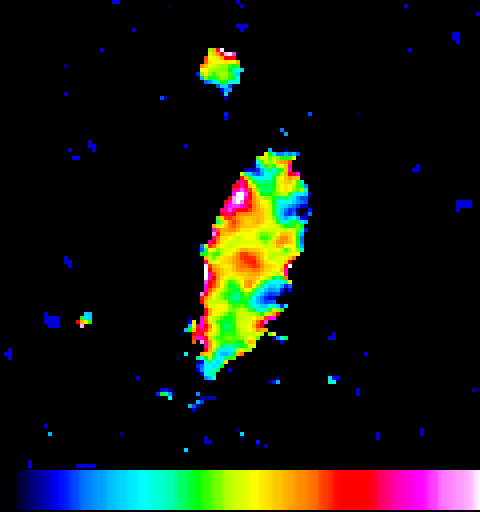}
\includegraphics[width=0.19\textwidth,frame]{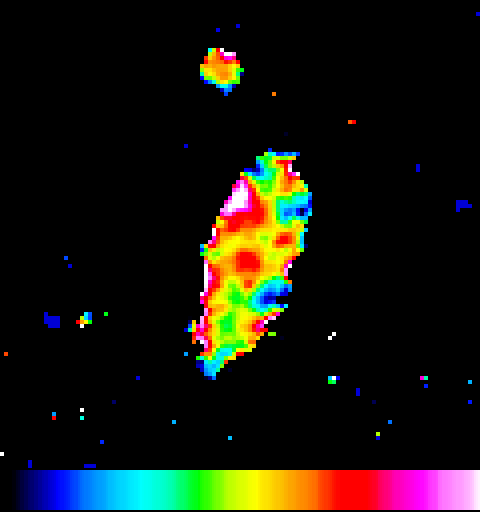}
\includegraphics[width=0.19\textwidth,frame]{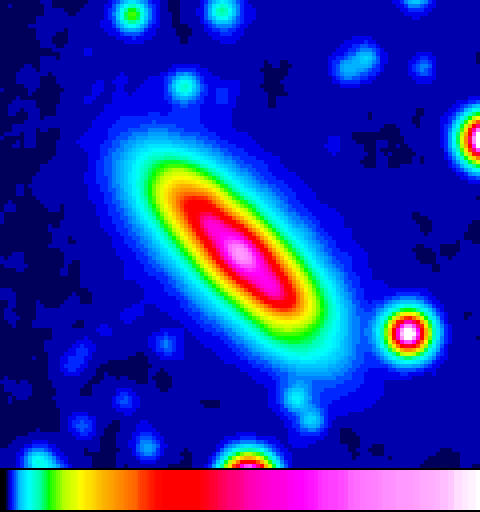}
\includegraphics[width=0.19\textwidth,frame]{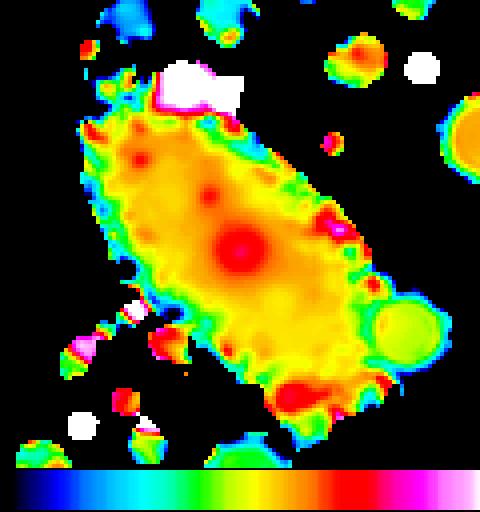}
\includegraphics[width=0.19\textwidth,frame]{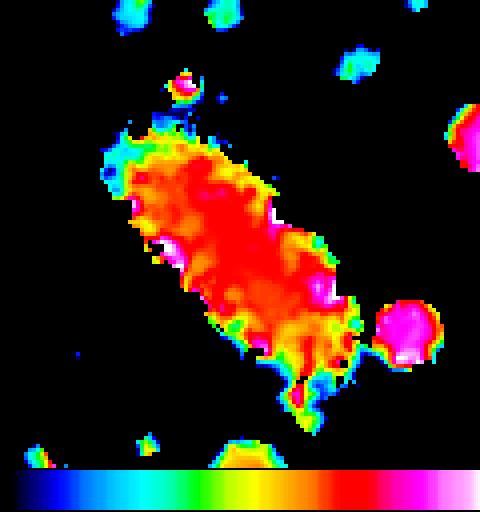}
\includegraphics[width=0.19\textwidth,frame]{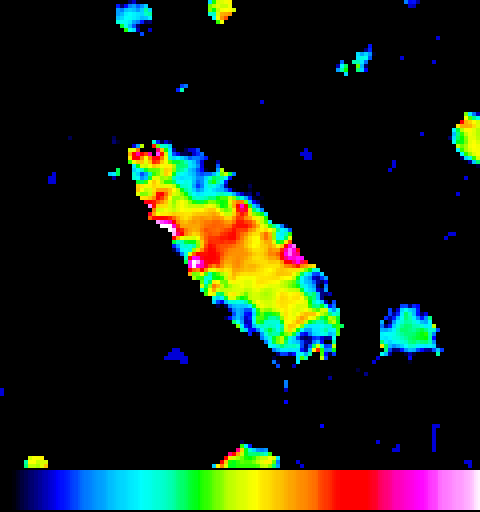}
\includegraphics[width=0.19\textwidth,frame]{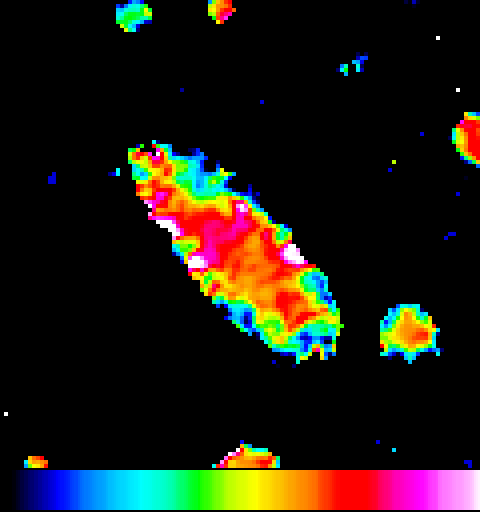}
\caption{Galaxies with red cores in $u-v$ (see Figure~\ref{gal_images_EpA} for details). 
Top: Sy-2 galaxy MCG+00-04-112 with a red core in $u-v$. Bottom: Sy-2 galaxy IC 1368 showing a narrow red structure in $u-v$ that is oriented along the minor axis and likely the ionisation cone of the central AGN revealed by the S7 Survey \citep{Dopita15}. 
}\label{gal_images3}
\end{center}
\end{figure*}

\subsection{Selecting further example galaxies from SkyMapper DR2}

In order to assess the significance of our change-index signal in the E+A galaxies, we would like to contrast them against galaxies with blue cores, caused by AGN or starbursts, and with cores of neutral $u-v$ colour. An exhaustive study of the entire galaxy population in the Southern hemisphere, i.e. of over 10 000 large nearby galaxies, would be interesting, but is beyond the scope of this paper. 

The photometric catalogues of DR2 are of limited value for the study of galaxies, because they lack forced-position photometry in consistent apertures. We do not plan to develop and execute such photometry on the image dataset of DR2 as part of this paper, and hence contend ourselves with a qualitative experiment at this stage: starting from the 2MASS Redshift Survey \citep[2MRS,][]{Huchra12} we analyse the catalogue photometry in small apertures in order to obtain a preliminary, albeit imprecise, view of the core properties in the nearby large-galaxy population. We then select a handful of galaxies with blue cores, average cores and red cores, for a qualitative analysis of their colour maps, as we did for the known E+A galaxies.

One problem of instantaneous photometry compared to forced-position photometry in consistent apertures is that bands of different sensitivity reveal different parts of a galaxy and lead to differing estimates of Petrosian radius and total flux. E.g., the low relative sensitivity in the SkyMapper $u$ and $v$ bands leads to a large underestimation of the Petrosian $uv$ fluxes, often by more than one magnitude, and hence the Petrosian colours from the \texttt{dr2.master} table are not helpful to study this sample of galaxies that have very large angular extent. 

We found it useful, however, to explore the colours of galaxy cores by considering measurements in smaller apertures that are listed in the \texttt{dr2.photometry} table. This photometry is not expected to be very precise for galaxies as large as those in the 2MRS catalogue, mostly due to background issues. Still, we found that the $5\arcsec$-aperture magnitudes, corrected for growth-curve losses, appear to rank the population reasonably well in terms of colour-magnitude, such that we can discriminate blue, average and red galaxy cores, and pick a few sample objects. We average the existing repeat measurements per band and determine formal errors from the scatter.

We consider the sample of 2MRS galaxies at $v<6000$~km s$^{-1}$ and $E(B-V)<0.1$ with Main Survey coverage in DR2, which contains about 1,300 galaxies with $M_K<-22.5$ \citep[or $\log M_*/{\rm M}_\odot >10$ assuming a K-band $M/L$ ratio of 0.6 following][]{McGaugh14}. In Figure~\ref{CMDs} we show colour-magnitude diagrams (CMDs) of these galaxies with $v-g$ and $u-v$ colour vs. $M_K$. Objects are colour-coded by 2MASS morphology: E/S0 galaxies are shown in red and account for half the sample; they mostly populate a red sequence, but have a tail to very blue $v-g$ core colours. Spirals of type Sa-Sc are shown in green, and the colours of their inner bulges are often consistent with the red sequence of E/S0 galaxies; however, they show a richer tail to blue colours. Finally, Sd and irregular galaxies are shown in blue and have the largest fraction of blue galaxy centres.

As expected, the $u-v$ CMD looks very different. Following our models in the previous section we expect galaxies with slow changes in star-formation rate to form a narrow ridge with $u-v \approx 0.35 + 0.3*E(B-V)$. Indeed, we find most E/S0 galaxies to scatter tightly around $0.39 \pm 0.057$ after clipping outliers. Owing to the rudimentary photometry, the colour errors are quite large, nearly on the order of the scatter suggesting that the true distribution is tighter. Several apparent outliers have very large errors, making them insignificant. From this diagram, we pick two galaxies each from three zones, with blue, average and red $u-v$ colours. The galaxies are marked with circles in the CMDs and include an S0 galaxy with a blue core.

\subsubsection{Example galaxies with blue cores}

Two galaxies with blue cores are shown in Figure~\ref{gal_images} and listed in Tables~\ref{sample} and \ref{sample2}. Their $v-g$ and $g-i$ colours are consistent with either starbursts or broad-line AGN (Sy-1 cores), but the $u-v$ colours of both are as extremely blue as is predicted only for AGN. A query to SIMBAD\footnote{ \url{http://simbad.u-strasbg.fr}} reveals their identify as the two Seyfert-1 galaxies Markarian 1044 and IC 1524. This result demonstrates that the SkyMapper filter set is capable of identifying perhaps all Southern Sy-1 galaxies at $z<0.03$, should any of them remain still unknown. 

In recent years, repeat spectroscopy of Seyfert-1 galaxies has revealed that unexpectedly many of them show variation in their appearance, changing between type 1 and type 2 on timescales of years or sometimes even months, through a (dis-)appearance of their broad emission-line region and their blue continuum \citep[``changing-look AGN'', e.g.][and references therein]{Shappee14}, but see also \citet{MacLeod16} and \citet{Yang18}. The phenomenon seems rare enough that many cases can only be found in large surveys, and the colours of galaxy cores measured with SkyMapper provide another snapshot revealing possible transient Seyfert-1 cores.

\subsubsection{Example galaxies with intermediate cores}

Two galaxies with intermediate cores are shown in Figure~\ref{gal_images2} and listed in Tables~\ref{sample} and \ref{sample2}. Their $v-g$ and $g-i$ colours show no strong colour gradients, although the second object, the known Seyfert-2 galaxy NGC 1320, shows two large red elongated features in $g-i$, which are likely large dust lanes as revealed by an HST image \citep{Ferruit00}. No strong trends are apparent in $u-v$ or the change index either. These examples serve as reference galaxies with no noticeable events.

\subsubsection{Example galaxies with red cores}

Two galaxies with red cores are shown in Figure~\ref{gal_images3} and listed in Tables~\ref{sample} and \ref{sample2}. Both are known Seyfert-2 galaxies, whose $g-i$ colours show colour gradients with red cores. While the first object shows also a pronounced red core in $v-g$ suggesting a core devoid of A stars, the second object, IC 1368, shows a uniformly red disk in $v-g$. In $u-v$ and the change index maps, the first shows a small red core, which is either due to recent quenching of star formation or strong OII emission from the Seyfert Narrow-Line Region. 

IC 1368 exhibits an extended red zone in $u-v$, oriented along the minor axis of the galaxy. IC 1368 was also observed by the Siding Spring Southern Seyfert Spectroscopic Snapshot Survey \citep[S7,][]{Dopita15}, which revealed an extended Narrow-Line Region (ENLR) with high OII line flux and high gas velocity dispersion that is similarly shaped and aligned as the $(u-v)$-red feature we see here. Thus, we can prominently see the ionisation cone from the central AGN in IC 1368.

\begin{figure*}
\begin{center}
\includegraphics[width=0.19\textwidth,frame]{title_g.png}
\includegraphics[width=0.19\textwidth,frame]{title_gmi.png}
\includegraphics[width=0.19\textwidth,frame]{title_vmg.png}
\includegraphics[width=0.19\textwidth,frame]{title_umv.png}
\includegraphics[width=0.19\textwidth,frame]{title_ci.png}
\includegraphics[width=0.19\textwidth,frame]{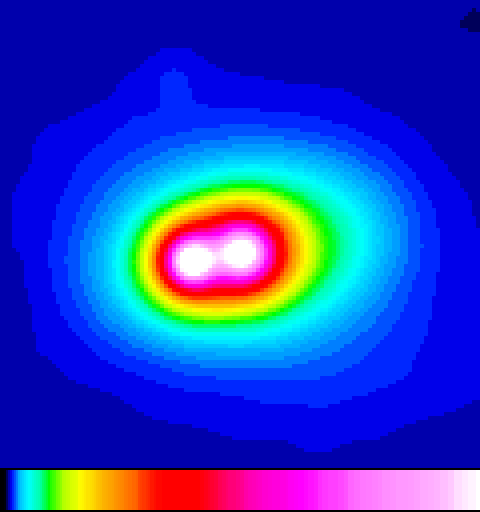}
\includegraphics[width=0.19\textwidth,frame]{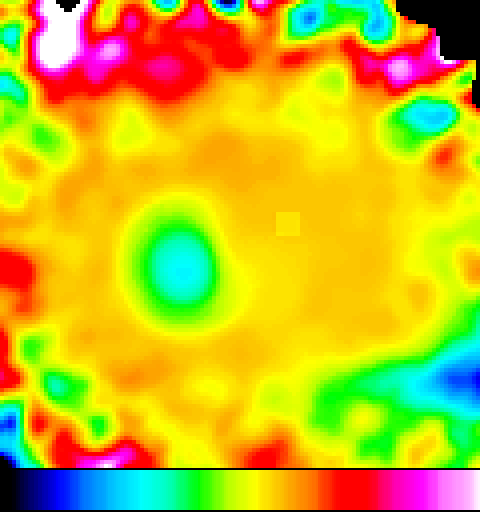}
\includegraphics[width=0.19\textwidth,frame]{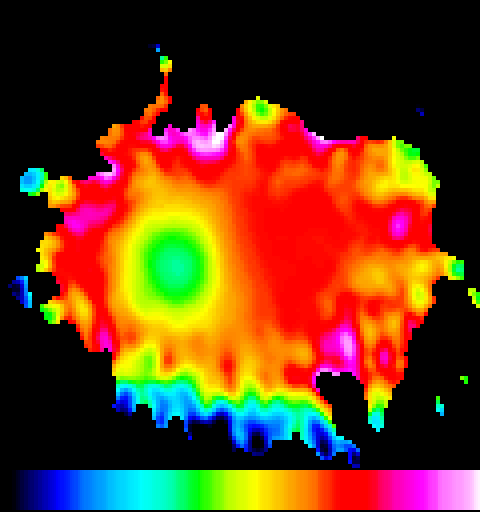}
\includegraphics[width=0.19\textwidth,frame]{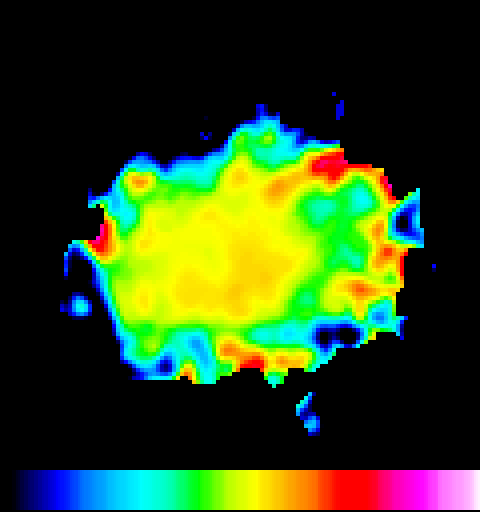}
\includegraphics[width=0.19\textwidth,frame]{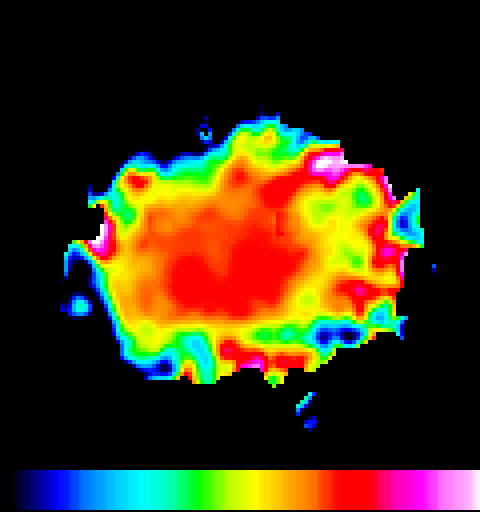}
\caption{Galaxy NGC 1321 (a.k.a. Mrk 608) with a blue star projected against it, which can be mistaken for a binary core. The star is blue in $g-i$ and $v-g$, neutral in $u-v$, but very red in the change index, which is typical for blue halo stars.
}\label{gal_images4}
\end{center}
\end{figure*}

\subsubsection{Chance projections with stars}

While investigating the objects above we noticed an apparently interesting companion galaxy: the Seyfert-2 galaxy NGC 1320 forms an interacting pair with NGC 1321. A tidal bridge of stellar light connecting the two is clearly visible in the SkyMapper $griz$ images. NGC 1321 appears at first sight to have a double nucleus, but it is little mentioned in the astronomical literature. 

We first followed the idea of a double nucleus, but struggled with interpreting the colours of the apparent, blue companion core, because it seemed neutral in $u-v$, but red in the change index. Only then did we find, that NGC 1321 is indeed a known chance projection of a lenticular galaxy with a bright blue halo star \citep{Petrosyan80}, and confirmed that the combination of colours is typical for those.




\section{Discussion}

Our results indicate that we can easily recognise E+A galaxies and Seyfert-1 galaxies from the core colours of galaxies, at least out to a distance of $\sim 100$~Mpc. Unfortunately, moderate nuclear activity is degenerate with changes in star-formation rate when no other information is available apart from SkyMapper colours. Outside of galaxy cores nuclear activity should not be much of an issue, so we should be able to recognise quenching and bursting phenomena across the face of galaxies once the over-subtraction issues of DR2 are addressed. However, there are large uncertainties for a quantitative analysis: the signal amplitude of the change index depends not only on the timescale of quenching, but also on metallicity. Also, faint stars can introduce noise into a colour map without being recognised. 

The change index derived here is not intended to be a powerful quantitative tool for studying complex star-formation histories of galaxies, which would require information of higher dimension to break various degeneracies. It is, however, a useful tool to point us to notable galaxies and probably still useful to highlight quenching phenomena outside of active galaxy cores. It is also a free tool that comes at no extra cost on top of the SkyMapper Southern Survey, which will provide change maps for all nearby Southern galaxies.

The main benefit of the new SkyMapper change index is that it responds to quenching activity within $\sim 20$~Myr and localises it within nearby galaxies across the sky. Quenching has been studied by many authors, but an-often cited result is that the galaxies identified as being quenched are caught too late in the process and it appears hard to reconstruct their earlier history. 

E.g., \citet{Scha07b} find that nearby galaxies with early-type morphology undergo instantaneous quenching, i.e. their star formation declines with $\tau_q < 250$~Myr, which is inconsistent with a simple scenario of gas exhaustion. Also, \citet{Wong12} find that nearby post-starburst galaxies show early-type morphology and hence argue that morphological transformation from presumably spiral progenitors must have been fast. 

Our change index should be able to see the onset of quenching earlier in the process and reveal galaxies before they reach the full post-starburst phase; it may assist in piecing together an empirical picture of the full transformation process. The SkyMapper Southern Survey is expected to provide this diagnostic for all galaxies in a local volume with 100~Mpc radius.

\citet{Scha07b} further argue that nearby galaxies of late-type morphology undergo slow quenching with $\tau_q<1$~Gyr, and semi-quenched galaxies that are seen now away from the star-forming main-sequence \citep[e.g.][and references therein]{Noeske07,Lee15} have undergone much change before the time of observation, making it hard to reconstruct the history of the progenitors. If this quenching time scale applies locally to all parts of a galaxy, none of them will appear remarkable in the light of our change index diagnostic.

However, while the global change in these galaxies is slow, local change could be faster; a locally fast effect that spreads spatially only slowly would integrate to a weak, diluted global signal of slow decline. Our maps of the change index would then see a local fast signal undiluted and at high contrast. This would help us to pinpoint the quenching phenomenon, while galaxies are still on the star-forming main-sequence and trace its spatial progression with a large sample of galaxies. 

If locally fast quenching exists in globally slowly quenched late-type galaxies \citep[e.g.][]{Fossati18}, then our change index maps will provide new clues to better understand how much quenching is related to halo mass, environment and group interactions. We will be able to see the interplay between galactic bars and quenching that appear to play a role in the secular evolution of late-type galaxies \citep{Masters11,Cheung13,Wang19}.

Finally, we should be able to differentiate whether dwarf galaxies undergo rapid or slow quenching when they interact with a hot halo and tidal field of their central galaxy \citep[e.g.][]{Wetzel15,Hausammann19}. Our change index could show the onset of episodes of rapid quenching and help constrain the role of the processes at work.

\section{SUMMARY}\label{summary}

We present a proof-of-concept for a bursting/quenching diagnostic that is purely based on galaxy images in the SkyMapper survey and will thus become available for the whole Southern sky. The new diagnostic is a pleasant by-product of a feature designed for other purposes. SkyMapper filters were designed for classifying stars in the Milky Way, and we realised that what is useful for characterising stars is also useful for characterising stellar populations in galaxies at redshift ``zero''. 

The SkyMapper change index is effective to a distance of at least 100 Mpc, while the spectra of more distant galaxies are redshifted out of sync with the spectral passbands. In practice, the diagnostic might be mostly useful as a quenching indicator, since high specific star-formation rates can already be observed directly in the form of UV-blue colours.

In this paper, we discussed a few example objects with characteristic signals in $u-v$ colour and change index maps, including E+A galaxies, Seyfert-1 galaxies and galaxies without notable signals. The $u$ band images in SkyMapper DR2 are affected by background over-subtraction, which precludes a full analysis across the face of galaxies, and motivated a restriction to signals from the galaxies' bright cores for this work. We expect that SkyMapper DR3 will fix this issue and allow a study of faint features on the outskirts of galaxies. 

Once complete, the SkyMapper Southern Survey will provide an inventory of Southern galaxies and the presented method allows us to create an SFR change index map of the entire Southern hemisphere with useful data on all Southern galaxies within 100~Mpc distance. This volume includes more than 10 000 galaxies with over 1~arcmin diameter, and a full analysis may reveal in detail the occurrence and spatial progression of changes in the birth rate of stars within the population of nearby galaxies. Interesting results could be expected especially in dwarf galaxies, interacting galaxies, compact groups, barred galaxies and include the discovery of new changing-look AGN.

\begin{acknowledgements}
This research was partly supported by the Australian Research Council Centre of Excellence for All-sky Astrophysics (CAASTRO), through project number CE110001020. JG acknowledges support from CAASTRO in contributing to this work.
The national facility capability for SkyMapper has been funded through ARC LIEF grant LE130100104 from the Australian Research Council, awarded to the University of Sydney, the Australian National University, Swinburne University of Technology, the University of Queensland, the University of Western Australia, the University of Melbourne, Curtin University of Technology, Monash University and the Australian Astronomical Observatory. SkyMapper is owned and operated by The Australian National University's Research School of Astronomy and Astrophysics. The survey data were processed and provided by the SkyMapper Team at ANU. The SkyMapper node of the All-Sky Virtual Observatory (ASVO) is hosted at the National Computational Infrastructure (NCI). Development and support of the SkyMapper node of the ASVO has been funded in part by Astronomy Australia Limited (AAL) and the Australian Government through the Commonwealth's Education Investment Fund (EIF) and National Collaborative Research Infrastructure Strategy (NCRIS), particularly the National eResearch Collaboration Tools and Resources (NeCTAR) and the Australian National Data Service Projects (ANDS).
We thank Mark Krumholz for advice on using his population synthesis code SLUG, and Adam Thomas for his insights into the S7 Survey.
This research has made use of the SIMBAD database, operated at CDS, Strasbourg, France.
\end{acknowledgements}

\bibliographystyle{pasa-mnras}

\end{document}